\documentclass[%
 reprint,
 amsmath,amssymb,
 aps,
 prx,
]{revtex4-2}

\usepackage{graphicx}
\usepackage{dcolumn}
\usepackage{bm}
\usepackage{braket}
\usepackage{xcolor}
\usepackage[normalem]{ulem}
\usepackage{amsmath}
\usepackage{bbm}
\usepackage{upgreek}

\begin{document}

\preprint{APS/123-QED}

\title{Ytterbium Nuclear-Spin Qubits in an Optical Tweezer Array}

\author{Alec Jenkins}
\thanks{These two authors contributed equally}
\author{Joanna W. Lis}
\thanks{These two authors contributed equally}
\author{Aruku Senoo}
\author{William F. McGrew}
\author{Adam M. Kaufman}
\email{adam.kaufman@colorado.edu}
\affiliation{JILA, University of Colorado and National Institute of Standards and Technology,
and Department of Physics, University of Colorado, Boulder, Colorado 80309, USA}%

\date{\today}

\begin{abstract}

We report on the realization of a fast, scalable, and high-fidelity qubit architecture, based on $^{171}$Yb atoms in an optical tweezer array. We demonstrate several attractive properties of this atom for its use as a building block of a quantum information processing platform. Its nuclear spin of 1/2 serves as a long-lived and coherent two-level system, while its rich, alkaline-earth-like electronic structure allows for low-entropy preparation, fast qubit control, and high-fidelity readout. We present a near-deterministic loading protocol, which allows us to fill a 10$\times$10 tweezer array with 92.73(8)\% efficiency and a single tweezer with 96.0(1.4)\% efficiency. In the future, this loading protocol will enable efficient and uniform loading of target arrays with high probability, an essential step in quantum simulation and information applications. Employing a robust optical approach, we perform submicrosecond qubit rotations and characterize their fidelity through randomized benchmarking, yielding 5.2(5)$\times 10^{-3}$ error per Clifford gate. For quantum memory applications, we measure the coherence of our qubits with $T_2^*$=3.7(4) s and $T_2$=7.9(4) s, many orders of magnitude longer than our qubit rotation pulses. We measure spin depolarization times on the order of tens of seconds and find that this can be increased to the 100 s scale through the application of a several-gauss magnetic field. Finally, we use 3D Raman-sideband cooling to bring the atoms near their motional ground state, which will be central to future implementations of two-qubit gates that benefit from low motional entropy.

\end{abstract}

\maketitle

\section{\label{sec:intro}Introduction}

The development of well-controlled, scalable qubit architectures is central to quantum science, and has seen rapid advances across a number of physical platforms~\cite{bruzewicz2019trapped,kjaergaard2020superconducting,kaufman2021quantum, sheng2018high, levine_parallel_2019, arute2019quantum, srinivas2021high, clark2021high, sung2021realization}. In this direction, neutral-atom qubits stored in optical arrays have made substantial progress in recent years~\cite{wang2016single, xia2015randomized, sheng2018high, levine_parallel_2019, graham_rydberg_2019, bluvstein2021quantum}. Combining the versatility of optical potentials with switchable Rydberg interactions creates a compelling platform for quantum information, simultaneously allowing dense, noninteracting qubit registers and two-qubit entangling operations~\cite{jaksch2000fast, schlosser2001sub, xia2015randomized,wang2016single, wilk2010entanglement,isenhower2010demonstration, levine_parallel_2019, graham_rydberg_2019, madjarov2020high, schine2021long, bluvstein2021quantum}. At the state of the art, large defect-free samples of hundreds of atomic qubits have been produced~\cite{ebadi2021quantum, scholl2021quantum}. Global single-qubit operations have reached below $10^{-4}$ error per gate~\cite{sheng2018high}, while errors at the $10^{-3}$ scale have been achieved in locally addressed arrays~\cite{xia2015randomized, wang2016single}. Two-qubit gates have reached errors at the several percent level~\cite{graham_rydberg_2019, levine_parallel_2019, yang2020cooling, schine2021long}, and have been employed in reconfigurable circuits~\cite{bluvstein2021quantum}. 

Though most neutral-atom quantum information experiments have focused on alkali atoms, a nascent thrust looks to extend optical tweezer technology beyond single-species alkali experiments. Pursuits with dual species, alkaline-earth atoms, and molecules~\cite{liu2018building,norcia2018microscopic, cooper_alkaline-earth_2018, saskin2019narrow,anderegg2019optical,singh2022dual} aim to translate the microscopic control of tweezers to new applications, as well as to harness new internal degrees of freedom for improved quantum science~\cite{ni2018dipolar, norcia_seconds-scale_2019, madjarov2020high, Anderegg2021Observation, barnes2021assembly}. In the case of alkaline-earth atoms,  marrying tweezer-based control with the long-lived optical transitions characteristic of this atomic group has enabled exploration of tweezer clocks~\cite{norcia_seconds-scale_2019, madjarov_atomic-array_2019, young_half_2020}. 
The rich internal structure of these atoms also offers new qubit modalities, ranging from Rydberg qubits~\cite{wilson2022trapping,madjarov2020high}, to optical-frequency qubits~\cite{schine2021long}, and to low-energy nuclear qubits~\cite{barnes2021assembly}. 

\begin{figure*}
\begin{centering}
\includegraphics{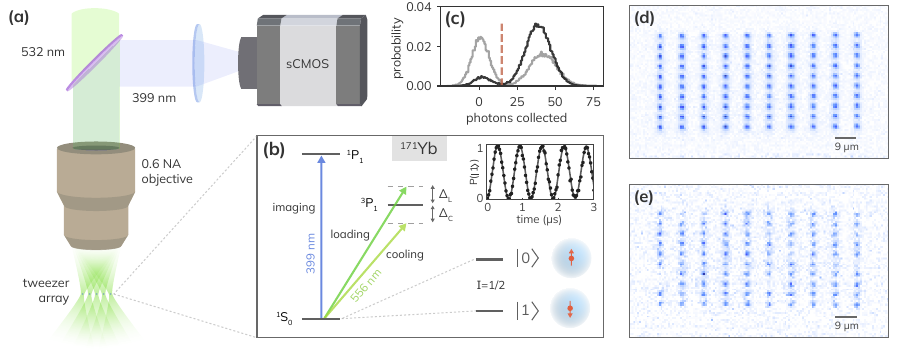}
\caption{\label{fig:apparatus_imaging} $^{171}$Yb optical tweezer arrays. (a) $^{171}$Yb atoms are trapped in an array of 100 optical tweezer sites. (b) The nuclear spin I=1/2 is an environmentally well-isolated two-level system. The level diagram shows the imaging, cooling, and enhanced loading transitions. Light scattered from the 399 nm $^1$S$_0 \rightarrow ^{1}$P$_1$ transition is collected with the high-NA objective. While imaging, we apply cooling beams red detuned by $\Delta_\mathrm{C}/(2\pi)=-2.04(5)$ MHz (11$\Gamma_\mathrm{g}$) from the light-shifted $^1$S$_0 \rightarrow ^{3}$P$_1$ $|F'=3/2,~m_{F'}=\pm1/2\rangle$ transition. The inset shows driven Rabi oscillations of the nuclear spin with submicrosecond $\pi$ times. (c) Histograms of collected photons over all tweezer sites for 120 ms exposures under two different loading schemes--- enhanced loading (black line) and standard (zero-field, red-detuned light) loading (gray line). The dashed red line is the threshold that we use to define detection of an atom in the enhanced loading case. (d) Average of 500 images of the tweezer array. (e) Single-shot image with near-deterministic tweezer array loading. 
}
\end{centering}
\end{figure*}

The nuclear spin of fermionic isotopes of alkaline-earth(-like) atoms provides a particularly attractive qubit. It has low (first-order) magnetic field sensitivity in states with zero electronic angular momentum, and it is expected to be robust to dephasing and decoherence mechanisms that arise in far-detuned optical traps due to a lack of hyperfine coupling~\cite{kuhr2005analysis, sheng2018high, dorscher2018lattice}. Moreover, it can be paired with long-lived electronic degrees of freedom for novel readout and two-qubit gate schemes~\cite{madjarov2020high, schine2021long}. Even beyond atom-based systems, nuclear-spin qubits have been the focus of efforts in the solid state, due to their decreased sensitivity to environmental perturbations~\cite{kane1998silicon,childress2006coherent,hensen2020silicon}. Very recent work with $^{87}$Sr nuclear-spin qubits in optical tweezers showed long coherence times ($T_2^* = 21(7)$ s) and single-site control; in this case, the methods needed to isolate a qubit from a large native nuclear spin (9/2) limits single-qubit gate times and can cause dissipation during gate operations~\cite{barnes2021assembly}. Further, standard qubit characterization methods---such as randomized benchmarking (RB)---have yet to be applied to this form of nuclear-spin qubit.

A compelling alternative candidate for a nuclear-spin-based qubit is $^{171}$Yb, which naturally is spin 1/2, the simplest nuclear-spin structure of any fermionic alkaline-earth(-like) atom (see Fig.~\ref{fig:apparatus_imaging}b) ~\cite{noguchi2011quantum, braverman2019near,mcgrew2018atomic}. Importantly, while prior work explored bosonic isotopes of ytterbium~\cite{saskin2019narrow}, this fermionic isotope has yet to be loaded, trapped, and manipulated in optical tweezers. Here, we report such methods, reveal scalability characteristics of $^{171}$Yb that are powerful for quantum science broadly, and benchmark techniques for controlling the nuclear qubit on submicrosecond timescales. 

Employing the narrow-line transitions, we show that $^{171}$Yb exhibits favorable properties for rapidly realizing large arrays that are both defect-free and at ultracold temperatures. The former fulfills a common need in quantum science to create large, uniformly filled qubit registers, while the latter aids single- and two-qubit gate fidelities as well as endeavors where motional-ground-state preparation is required~\cite{kaufman2014Two, cairncross2021Assembly, levine_parallel_2019, schine2021long}. To overcome the stochastic nature of the single-atom loading process into tweezers, low-defect samples have been prepared  either through active rearrangement or tailored collisions for enhanced loading~\cite{schlosser2001sub, grunzweig2010near, lester2015rapid, endres2016atom,barredo2016atom, brown2019gray}. Inspired by recent advances~\cite{brown2019gray}, we use a narrow-line cooling transition to achieve near-deterministic loading of a $10\times10$ array of $^{171}\textrm{Yb}$ atoms with single-site occupancy of 92.73(8)\% and 96.0(1.4)\% for a single tweezer. At the same time, we exploit the nuclear spin for Raman-sideband cooling to reach near ground-state temperatures ($n_r = 0.14(3),~n_z = 0.13(4)$), which will aid future manipulation of the clock transition and high fidelity Rydberg-mediated two-qubit gates~\cite{monroe1995resolved, kaufman2012cooling, thompson2013coherence, de2018analysis,schine2021long}. These results are of central importance to the rapid generation of large, uniformly filled arrays of high fidelity qubit registers, quantum simulation experiments based on low-entropy spin models, and optical atomic clocks in tweezers and lattices using alkaline-earth atoms~\cite{de2018analysis,levine_parallel_2019,young_half_2020,oelker2019demonstration,scholl2021quantum,ebadi2021quantum}. 

With nearly 100-atom arrays, we demonstrate universal single-qubit control of the nuclear-spin qubit with submicrosecond operations. The state of the art in single-qubit fidelities with neutral atoms has been achieved with microwave-driven transitions with pulse times of several tens of microseconds~\cite{wang2016single,xia2015randomized,sheng2018high}. With the potential for submicrosecond two-qubit gates through Rydberg-mediated interactions~\cite{graham_rydberg_2019, levine_parallel_2019, madjarov2020high, schine2021long}, reaching high-fidelity single-qubit gates on similar timescales is important for fully realizing the potential computational speeds of a neutral-atom system. Enabled by the low-energy nuclear qubit, we use a robust rotation scheme based on single-beam Raman transitions to reach 170 ns $\pi/2$ pulses, which is readily extendable to rapid single-qubit addressing. In a globally illuminated array, we reach a Clifford gate error of $5.2(5) \times 10^{-3}$, characterized through randomized benchmarking. We establish a straightforward path to errors of $\ll 10^{-4}$---a commonly held scale for resource efficient error correction~\cite{terhal2015quantum, fowler2012surface}---based on our characterization of the trapped qubit coherence time ($T_2^* = 3.7(4)~\textrm{s}, T_2 = 7.9(4)~\textrm{s}$) and quantified error sources. 

\section{\label{sec:apparatus}Preparation and detection}
We generate arrays of optical tweezers with 532 nm laser light using a pair of acousto-optic deflectors (AODs) crossed at 90$^\circ$ and a 0.6 NA microscope objective. To reach favorable conditions for loading the tweezers, the atoms are first captured in a 3D magneto-optical trap (MOT) operating on the broad, $\Gamma_\mathrm{b}/(2\pi)=29$ MHz, 399 nm $^1$P$_1$ transition, loaded from a 2D MOT~\cite{nosske2017two, dorscher2013creation}. The atoms are then transferred to a narrow MOT that uses the 556 nm ${^3}\mathrm{P}_1$ transition with linewidth $\Gamma_\mathrm{g}/(2\pi)=183$ kHz. From the narrow green MOT, atoms are loaded into the tweezer array and we isolate single atoms in the tweezers with a beam blue detuned from one of the ${^3}\mathrm{P}_1$ hyperfine states.

Figure \ref{fig:apparatus_imaging} gives an overview of $^{171}$Yb trapping and imaging. Unlike the $^{174}$Yb isotope \cite{yamamoto2016ytterbium, saskin2019narrow}, the ${^1}\mathrm{S}_0 \leftrightarrow{}{^3}\mathrm{P}_1$ transition of $^{171}$Yb is not near-magic at 532 nm, and a magic angle does not exist at this trapping wavelength for any orientation of the quantization axis~\cite{norcia2018microscopic}. For nonmagic tweezers, small variations in trap depths can give large variations in scattering rates from the narrow ${^3}\mathrm{P}_1$ levels due to the resulting variations in detuning from resonance. This effect leads to low imaging fidelities when detecting fluorescence scattered from ${^3}\mathrm{P}_1$. For this reason, we instead detect the presence of atoms in the tweezer array by scattering light off the broader ${^1}\mathrm{P}_1$ transition while simultaneously cooling with the narrower green MOT beams, detuned by $\Delta_\mathrm{C}/(2\pi)=-2.04(5)$ MHz, or $11\Gamma_\mathrm{g}$, from the light-shifted ${^3}\mathrm{P}_1$ $|F'=3/2,m_{F'}=\pm1/2\rangle$ resonance. For $^{171}$Yb, this method produces a narrower atom peak in the photon collection histogram (Fig.~\ref{fig:apparatus_imaging}c) and allows us to set a count threshold for detecting the presence of an atom with higher fidelity than when detecting scattered light from ${^3}\mathrm{P}_1$. However, the ${^1}\mathrm{P}_1$ scattering rates are limited by the ${^3}\mathrm{P}_1$ cooling rates, requiring us to extend the imaging times in order to minimize atom loss \cite{saskin2019narrow}. 

The scattered 399 nm light is collected by our microscope objective and imaged onto a scientific CMOS (SCMOS) camera (Fig. \ref{fig:apparatus_imaging}d,e). We assess detection performance using histograms of collected photons per tweezer site (Fig. \ref{fig:apparatus_imaging}c). From these data, we calculate the average detection infidelity as the probability of misidentifying the presence of an atom averaged over the cases with and without an atom \cite{norcia2018microscopic}. The imaging duration is typically 120 ms and the resulting infidelity is around 0.3\%. The loss probability is calculated as the fraction of tweezer sites in which an atom is identified in the first image but not the second with a small correction accounting for the imaging infidelity. The infidelity-corrected loss of 2.51(1)\% from the dataset shown in Fig. \ref{fig:apparatus_imaging}c is typical of the imaging loss in all experiments. For careful calibration of cooling parameters, we find it is possible to use imaging durations as short as 60 ms with similar loss rates and $0.6\%$ imaging infidelity. In the future, using a better-suited camera technology  (electron multiplying CCD [EMCCD]), we expect to achieve this scale of imaging infidelity and loss $\leq 1\%$ for an imaging time of 25 ms. (See Appendix \ref{subsec:image} for more details.)

\section{\label{sec:enhanced_loading} Near-deterministic loading}

Reliable assembly of large defect-free patterns of qubits is a major pursuit for the neutral-atom tweezer array platform. While the number of atoms loaded from the MOT into a tweezer follows a Poisson distribution, isolating single atoms often employs light-assisted collisions which map even and odd atom numbers into 0 and 1 respectively. The resultant stochastic loading pattern can be rearranged into a defect-free array with movable tweezers that drag atoms into the desired locations~\cite{barredo2016atom, endres2016atom, schymik2020enhanced}. However, the required time and success probability of reaching a defect-free array scales adversely with the number of atoms to be moved~\cite{barredo2016atom, endres2016atom, schymik2020enhanced}. Additionally, at $\sim$50\% loading efficiency the required optical power to produce uniform arrays of a given size is effectively doubled. The development of (near-)deterministic loading protocols can aid the process of rearrangement and alleviate the task of scaling the system size. A prominent example, so far demonstrated only for alkali atoms, is a protocol based on blue-shielded collisions~\cite{grunzweig2010near, lester2015rapid,brown2019gray}. At the state of the art, a combination of gray-molasses cooling and repulsive molecular potential leads to $\sim$80\% loading efficiencies in 10$\times$10 tweezer arrays and up to 90\% for a single tweezer~\cite{brown2019gray}. An outstanding question has been whether the narrow lines of alkaline-earth atoms could be used for similar enhanced loading schemes~\cite{norcia2018microscopic,cooper_alkaline-earth_2018,saskin2019narrow}. Here, we demonstrate the first realization of such a protocol.

\begin{figure}
\includegraphics{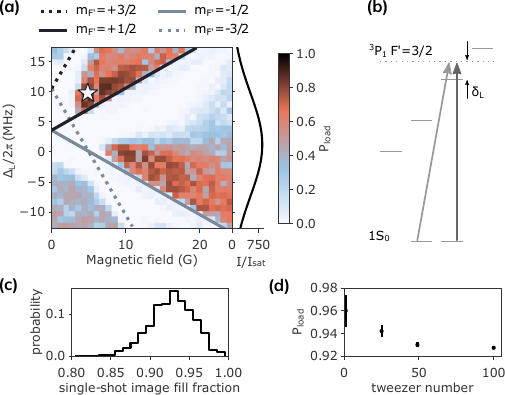}
\caption{\label{fig:loading} Near-deterministic loading. (a) Loading efficiency ($\mathrm{P_{load}}$) dependence on applied magnetic field and loading beams detuning ($\Delta_L$) from the free-space resonance. The star denotes the parameters chosen for daily operation: magnetic field of 4.9 G and $\Delta_\mathrm{L}/2\pi= +9.76$ MHz. The intensity of the loading beams changes as a function of $\Delta_\mathrm{L}$, in the starred region corresponding to 515$I_\mathrm{sat}$, where $I_\mathrm{sat} = 0.138$ mW/cm$^2$ is the saturation intensity of the ${^1}\mathrm{S}_0\leftrightarrow {^3}\mathrm{P}_1$ transition. The resonance frequencies of the transitions to the $m_{F'}$ states are plotted as a function of the magnetic field. (b) Levels involved in the loading scheme for parameter space denoted with the star in (a). Accounting for tweezer light and magnetic shifts, the loading beams are blue detuned from $m_{F'}=+1/2$ by $\delta_\mathrm{L} = + 2\pi \times 2.8$ MHz (15.6 $\Gamma$). (c) Probability of obtaining a given fill fraction for a single-shot image. (d) Optimal average loading efficiency ($\mathrm{P_{load}}$) as a function of the tweezer array size. The error bars correspond to standard deviations of the binomial distributions given by the measured probabilities.}
\end{figure}

We start by preparing a compressed narrow-line MOT to increase the atom density and thus ensure that each tweezer is filled with at least one atom on average. The atoms are loaded into  $17.1(5)$-MHz-deep tweezer traps (6 mW/tweezer) and addressed with three orthogonal pairs of circularly polarized counterpropagating beams tuned near the $F'=3/2$ hyperfine levels of ${^3}\mathrm{P}_1$. The tweezer light induces differential light shifts of $3.75(16)$ and $10.22(18)$ MHz on the $|m_{F'}|=1/2$ and $|m_{F'}|=3/2$ states, respectively, as compared to ${^1}\mathrm{S}_0$. We further split the $m_{F'}$ levels with a magnetic field applied along the tweezer polarization direction. Figure \ref{fig:loading}a shows two regions of enhanced loading ($>50\%$), observed for loading beams blue detuned of the transitions to the $m_{F'}=-1/2$ and $m_{F'}=+1/2$ states. The magnetic field and detuning chosen for daily operations are denoted with a star. At these conditions, the loading beams are blue detuned from $m_{F'}=+1/2$ by $\delta_L = +2\pi \times 2.8$ MHz (15.6 $\Gamma$) (Fig. \ref{fig:loading}b). The atoms are addressed with the beams of total intensity of 515$I_\mathrm{sat}$ for 35 ms, where $I_\mathrm{sat} = 0.138$ mW/cm$^2$ is the saturation intensity of the ${^1}\mathrm{S}_0\leftrightarrow {^3}\mathrm{P}_1$ transition. With these parameters, single atoms are loaded into the tweezers with 92.73(8)\% probability averaged over the 10$\times$10 array (Fig. \ref{fig:loading}c). We also investigate loading efficiency as a function of the tweezer array size. We observe consistent loading of the array at $>$90\% efficiency for tweezer numbers ranging from one to 100 traps (Fig. \ref{fig:loading}d). Maximal loading efficiency increases for smaller trap numbers, similar to results reported for alkali atoms \cite{brown2019gray}. In particular, for a single tweezer, we observe loading efficiencies as high as 96.0(1.4)\%, which suggests that a limitation in the larger arrays is the initial loaded atom number. Hence, we hypothesize that the single tweezer loading performance should be accessible in larger arrays by improving the initial cloud density, for instance, by use of a reservoir trap~\cite{anderegg2019optical}. Additionally, we investigate deterministic loading in shallower tweezers. For a 7$\times$7 array and half the usual tweezer depth ($\sim$8.5 MHz,$\sim$3 mW/trap), we reach 91.0(2)\% loading efficiency, further emphasizing the atom-scaling characteristics of this approach. 

The enhanced loading process arises from an interplay between cooling and light-assisted collisions with controlled energy transfer, which occurs for blue-detuned light~\cite{grunzweig2010near,lester2015rapid,brown2019gray}. After enhanced loading, we perform release-recapture experiments \cite{tuchendler2008energy} and measure temperatures of 5 $\upmu$K. This is to be compared with 12 $\upmu$K temperatures achieved through the cooling mechanism employed during imaging. These observations are consistent with a gray-molasses cooling effect observed in alkalis~\cite{brown2019gray}, and also reported in $^{173}$Yb isotope \cite{ono2021two}. In this case, by splitting $m_{F'}$ sublevels by more than $\Gamma_g$, the magnetic field plays a dual role: it isolates a three-level system where a gray-molasses mechanism can take place, and ensures that as the atom moves within the trap, it is not brought to resonance with other sublevels through the light shifts. At these blue detunings for which we observe optimal loading, we are also detuned by substantially less than the trap depth. In the usual blue-shielded loading picture~\cite{grunzweig2010near,lester2015rapid}, this suggests that the energy imparted per light-assisted collision is much less than the trap depth, so that the likelihood of both atoms remaining in the trap after a collision is high, while a small probability exists for one atom to leave, and an even smaller chance for both. We hypothesize then that the enhanced loading is the result of many collisions, where each collision is more likely to reduce the atom number by one atom at a time rather than in pairs. This enhances the probability for a single atom to remain at the end of the process. A quantitative model will be the subject of future investigations, as we expect a complicated cooperation between collisions, light shifts, load rates, and three-dimensional cooling dynamics to be responsible for the observed loading efficiencies. 

\section{\label{sec:qubit_rotations}Nuclear qubit control}

\begin{figure*}
\begin{centering}
\includegraphics{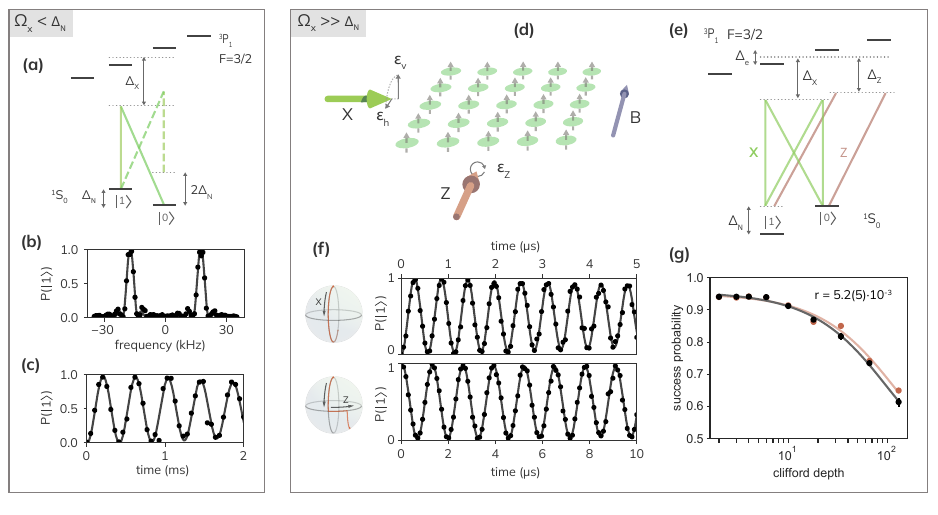}
\caption{\label{fig:qubit_rotations} Nuclear spin control. (a) The Raman level diagram corresponding to the weak-drive regime. Two copropagating beams drive Raman transitions between the nuclear-spin states. The two beams have variable relative frequency, one with $\pi$ polarization, and the other with polarization normal to the atom plane (both $\sigma_+$ and $\sigma_-$ polarizations). The rotations are performed through two excited states $\ket{F'=3/2,m_{F'} = \pm 1/2}$ but in the weak-driving regime one of these Raman transitions (dashed lines) is far off resonance compared to $\Omega_X$ and effectively does not contribute. The detuning from the $F'=1/2$ states is much larger than $\Delta_X$ and these states do not significantly affect the dynamics. (b) As we sweep the detuning between the Raman beams, the resonances corresponding to the two Raman pathways become visible. (c) Fixing the frequency to the left resonance, we drive Rabi oscillations at the kilohertz scale. The plotted probabilities of detecting the $\ket{1}$ state are normalized by the measured atom survival probability for the experiments in the absence of blow-away, which is 96.5(2)\% for both (b) and (c), see Appendix \ref{sec:SPAM}. (d) In the strong-driving regime, a single $X$ beam with polarization components along both the quantization axis and the atom plane normal drives two Raman transitions between the spin states. A second beam, $Z$, splits the nuclear-spin states and drives oscillations around the $Z$ axis. A small magnetic field $\mathbf{B}$ splits the nuclear-spin states by $1.25$ kHz. (e) Level diagram for $X$ and $Z$ beams. The beam detunings are $\Delta_X/(2\pi)\simeq -180$ MHz and $\Delta_Z/(2\pi)=-164$ MHz, while the excited state splittings are much smaller $\Delta_e/(2\pi)\simeq 2$ MHz. (f) The top panel shows $X$ Rabi oscillations at $1.77$ MHz. In the bottom panel, we measure $Z$ oscillations at $0.77$ MHz with a Ramsey-type sequence, where the $Z$ beam is turned on for variable time between two $X(\pi/2)$ pulses. The survival probability for these experiments without blow-away is 96.3(1)\%. The accompanying Bloch spheres show example trajectories for the two types of experiments. (g) Clifford randomized benchmarking using Clifford gates compiled of $X(\pi/2)$ and $Z(\pi/2)$. The target output state is randomized between $\ket{0}$ and $\ket{1}$. The black circles show the measured probabilities of obtaining the target state at a given Clifford gate depth. At each depth, we run 40 different sets of randomized experiments. The error bars are given by the standard deviation over the 40 sets of experiments. The extracted average gate infidelity per Clifford gate is $5.2(5)\times 10^{-3}$. The red circles show the simulated success probabilities using the estimated scattering error rates of the $X$ and $Z$ beams and shot-to-shot fractional intensity noise of 1\%, with 3\% fractional intensity noise from experiment to experiment (Appendix \ref{sec:qubit}). }
\end{centering}
\end{figure*}

We initialize and detect the ${^1}\mathrm{S}_0$ nuclear-spin states $\ket{m_F=+1/2}\equiv\ket{0}$ and $\ket{m_F=-1/2}\equiv\ket{1}$ using state-selective transitions to the ${^3}\mathrm{P}_1$ levels. Driving the $\sigma_+$ transition from $\ket{1}$ to the ${^3}\mathrm{P}_1$ $\ket{F'=1/2,~m_{F'}=+1/2}$ state pumps the ground ${^1}\mathrm{S}_0$ state spins to the $\ket{0}$ state. For spin detection, we apply the same beam, but frequency shifted to address the transition $\ket{0} \leftrightarrow \ket{F'=3/2,~m_{F'}=+3/2}$. This heats atoms in $\ket{0}$ out of the tweezers and atom survival in a subsequent image indicates the spin state. In order to address both of these transitions with the same beam, we use a fiber electro-optic modulator to produce a sideband at the hyperfine splitting of ${^3}\mathrm{P}_1$. The offset frequency is chosen to set the $\ket{F'=1/2,~m_{F'}=\pm1/2}$ resonances far from the $\ket{F'=3/2,~m_{F'}=+3/2}$ blow-away resonance of the carrier (19 MHz), but the finite separation does result in some optical pumping of the spin state for long blow-away times. The ideal blow-away time is set by a trade-off between unwanted pumping from $\ket{1}$ to $\ket{0}$ at long times and partial blow-away of the $\ket{0}$ state at short times. We minimize the combined detection infidelity resulting from this blow-away pulse, defined as the probability that an atom in $\ket{1}$ is pumped to $\ket{0}$ plus the probability that an atom in the $\ket{0}$ state survives blow-away. The minimum combined detection infidelity is typically around $6\times 10^{-3}$, and occurs for blow-away times around 5 ms. We measure the atom survival after optical pumping and blow-away detection, and taking into account the blow-away and imaging infidelities, estimate the preparation fidelity of the $\ket{0}$ state to be 0.996(1).

Nuclear-spin rotations are performed in two different regimes of driving strength. Figure \ref{fig:qubit_rotations}a shows a level diagram relevant to the ``low-Rabi regime," with a Rabi frequency $\Omega_X$ smaller than the nuclear-spin splitting $\Delta_N$. Two copropagating beams, one with $\pi$ polarization and the other with polarization normal to the quantization axis ($\sigma_+/\sigma_-$), are used to drive qubit Rabi oscillations. Two separated Raman resonances are observed as the frequency of one of the beams is varied (Fig. \ref{fig:qubit_rotations}b). The applied field here is $18$ G. In the low-Rabi regime, the Rabi oscillations are dominated by the single resonant pathway and the Rabi frequency does not depend on the relative phase of the $\sigma$ and $\pi$ polarization components. In this case, the speed of operations is limited by the splitting of the nuclear-spin states, typically small for achievable fields, $\gamma_{n,171}/(2\pi) = 751$ Hz/G \cite{gossard1964ytterbium}. 

To circumvent this limitation on speed, we explore whether it is also possible to perform high-fidelity operations in the opposite regime of $\Omega_X \gg \Delta_N$. Unlike when addressing one of the Raman resonances with a weak drive, in the high-Rabi case the Rabi frequency depends on the relative phase of the $\sigma$ and $\pi$ components. For an intermediate state detuning ($\Delta_{X}$) much greater than splittings of the excited state levels $\Delta_{X} \gg \Delta_e$, and relative phase $\phi$ between the $\pi$ and $\sigma$ single-photon Rabi frequency components, the Raman Rabi frequency is given by $\Omega_X \simeq \Omega_\pi \Omega_\sigma \text{cos}(\phi)/\Delta_X$, where the ratio of the coupling strengths $\Omega_\pi$ and $\Omega_\sigma$ is set by the ratio of $E_z$ and $E_y$ (Appendix \ref{sec:qubit}). The magnitudes of the two components $|\Omega_{\pm\sigma}|\equiv\Omega_\sigma$ are equal for our beam geometry. In this regime, we use a single beam (Fig.~\ref{fig:qubit_rotations}d) to drive Raman transitions simultaneously through the two separate pathways shown in Fig. \ref{fig:qubit_rotations}e, albeit with a small detuning given by the qubit splitting. Using a single beam has the advantageous feature that the phase between different polarization components is fixed. The phase that minimizes the ratio of qubit error rates to Raman Rabi frequency corresponds to a circularly polarized beam, but we note that here our beam is elliptically polarized.

Figure \ref{fig:qubit_rotations}f (top) shows qubit Rabi oscillations at $1.77$ MHz using the described approach. The applied field is 1.66 G, defining the quantization axis and splitting the nuclear-spin states by $\Delta_N/(2\pi) \equiv f_{0} - f_{1} = - 1.25$ kHz in the absence of the $X$ beam. When the $X$ beam is turned on, light shifts from the excited states cause a splitting in the opposite direction. For the Rabi frequency $\Omega_X/(2\pi) = 1.47$ MHz, used for $X(\pi/2)$ pulses below, the total splitting is estimated to be $\Delta_N/(2\pi) =  +54.2$ kHz. To perform arbitrary rotations of the spins, we need the ability to rotate about an axis different than that defined by our $X$ beam. This would normally be accomplished by varying the relative phase of two Raman beams, but in the high-Rabi case the coupling strength is phase dependent, becoming very small for a relative phase of $\pi/2$. This is the reason for including a second $Z$ beam that splits, but does not couple, the two nuclear-spin states (Fig. \ref{fig:qubit_rotations}d). Figure \ref{fig:qubit_rotations}f (bottom) shows oscillations at $0.77$ MHz about the $Z$ axis, measured using a Ramsey-type sequence: after a $X(\pi/2)$ rotation, the $Z$ beam is turned on for variable time, before a final $X(\pi/2)$ pulse is applied. This oscillation frequency corresponds to the splitting of the $\ket{0}$ and $\ket{1}$ states induced by differential light shift of the $Z$ beam, and is much larger than the $1.25$ kHz Zeeman field splitting observed in a Ramsey measurement without the $Z$ beam. Our typical $Z(\pi/2)$ pulses last 350 ns, corresponding to oscillations at $0.71$ MHz. 

We characterize the fidelity of nuclear qubit rotations using Clifford randomized benchmarking \cite{magesan2011scalable,olmschenk2010randomized, xia2015randomized}. We select $X(\pi/2)$ and $Z(\pi/2)$ rotations for characterization since we can compose the entire Clifford gate set from these two gates. With the software package pyGSTio \cite{pygsti}, we compile our $\pi/2$ gates into a set of Clifford RB experiments. The target output states are randomized between $\ket{0}$ and $\ket{1}$ and the probability of successfully measuring the target state is obtained over many repetitions of the experiment. Randomizing the target output gives a decay of fidelity that approaches 0.5 in the single-qubit case. Measurement infidelities and atom loss would give a different asymptote when using only one target state for every experiment, resulting in systematic bias in a fixed asymptote fit to the decay curve. The black circles in Fig. \ref{fig:qubit_rotations}g show the result of the randomized benchmarking up to a depth of 130 Clifford gates. We fit the success probability with a fixed-asymptote decay function $P_s(l)=0.5 + b\times p^l$, where $l$ is the gate depth, $p$ is the decay constant, and $b$ is a free parameter that accounts for a nonunity success probability at depth zero. From this fit, we extract the mean average gate infidelity $r=(1-p)/2=5.2(5)\times 10^{-3}$. A free-asymptote fit similarly gives $r=5(2)\times10^{-3}$. On average, there are 3.5 of the $X(\pi/2)$ and $Z(\pi/2)$ gates per single Clifford gate in these experiments. This means that the error per single $X$ or $Z$ gate is smaller than the error given for a single Clifford gate, although directly dividing the Clifford error by the average gate number likely underestimates the base $X (\pi/2)$ and $Z (\pi/2)$ gate errors. 

To understand the source of these gate errors, we measure and calculate the error rates due to intensity noise, scattering (both Raman and Rayleigh) of the qubit beams, and the detuning of the $X$ drive. We estimate the decoherence probability of an equal spin superposition due to the $X$ beam of $1.0\times 10^{-3}$ in the time of a single $X(\pi/2)$ pulse \cite{uys2010decoherence}. For the $Z$ beam, we find an equal superposition decoherence probability of $9\times 10^{-4}$ in the time of a single $Z(\pi/2)$ pulse. For intensity noise and detuning errors we expect error rates to be in the range $1\times 10^{-3}$ to $2\times 10^{-3}$ during a single $\pi/2$ gate (Appendix \ref{sec:qubit}). Using the (state-dependent) scattering error rates of the two beams, the measured fractional intensity noise (1\% pulse to pulse, and 3\% experiment to experiment), and the estimated detuning $\Delta_N$, we simulate the success probabilities for the exact sequence of $X$ and $Z$ gates applied in the measurements. The red circles in Fig. \ref{fig:qubit_rotations}g give the simulation results with the survival probability in the simulation scaled to match the measured value at a Clifford gate depth of zero. 

The good agreement between the measured decay and the decay due to the simulated errors indicates that these are likely the dominant sources of gate errors. The identified errors can be improved substantially (Appendix \ref{sec:qubit}), but scattering of the Raman beams and available laser power sets an important limit to the $X$ gate fidelity. For example, fixing the Rabi rate above, using 1 W of optical power in a $1\times0.02$ mm beam, and the ideal $X$ beam polarization, the scattering-limited error rate in a single $\pi/2$ pulse is $3\times 10^{-6}$ at +13 GHz detuning from $F'=3/2$. In the future, the single-beam Raman rotations on ${^1}\mathrm{S}_0\leftrightarrow{}{^3}\mathrm{P}_1$ could be paired with the individual atom addressing through the high-NA objective, for rapid local single-qubit gates with similar or even lower Clifford gate errors requiring significantly lower power. Moreover, for experiments with Rydberg interactions generated between atoms in the ${^3}\mathrm{P}_0$ level, it could be advantageous to perform similar nuclear-spin rotations on atoms in the ${^3}\mathrm{P}_0$ level directly. In that case, it would be possible to use the analogous transitions ${^3}\mathrm{P}_0\leftrightarrow{}{^3}\mathrm{D}_1$ or ${^3}\mathrm{P}_0\leftrightarrow{}{^3}\mathrm{S}_1$.

\section{\label{sec:coherence} Coherence characterization}

The coherence time of the nuclear qubit states sets one limit on the depth of operations that may be performed for algorithms or sensing measurements that utilize the nuclear-spin qubits. In this work, the number of coherent operations we can perform is limited almost entirely by other sources such as scattering and intensity noise. However, for future experiments, if slower processes are involved  (e.g. clock-state spectroscopy or atom moves~\cite{bluvstein2021quantum}) the nuclear-spin states can act as a quantum memory with long coherence times; further, if computational time is less of a priority, radio-frequency driving of the nuclear qubit can mitigate the dominant error sources observed with the Raman approach. In this section, we present coherence characterization of our qubit through Ramsey-type measurements. 

\begin{figure*}
\centering
\includegraphics{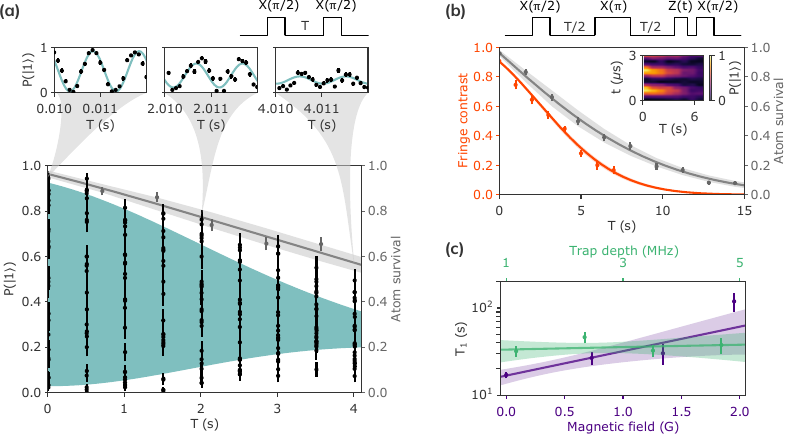}
\caption{\label{fig:coherence} Qubit coherence. (a) Ramsey experiment. Population of state $|1\rangle$ oscillates at the frequency given by the splitting of two ground states. The fit (teal) is a cosine of a single frequency and phase, with an envelope given by the lifetime and Gaussian dephasing. The green region in the lower plot corresponds to the envelope of the fit shown in the callouts. The $T_2^*$ extracted from the fit to the oscillations is 3.7(4) s. The measurement of atom lifetime in tweezers is also plotted (gray), with 1/e time of 6.42(3) s. The error bars correspond to standard deviations of the binomial distributions given by the measured probabilities. (b) Spin-echo experiment. Inset: population of state $|1\rangle$ oscillates with the duration of $Z(t)$ gate applied before the final $X(\pi/2)$ pulse. The contrast of the recorded fringe (orange) decreases with dark time $T$ due to the finite lifetime in tweezers (gray), of 7.13($^{+6}_{-5}$) s, and decoherence described by a Gaussian decay. The $T_2$ inferred from the fit is 7.9(4) s. The orange error bars are given by the square root of the covariance matrix diagonal entry corresponding to the fit contrast parameter. (c) Depolarization time. $T_1$ dependence on applied magnetic field (purple) and tweezer depth (green). $T_1$ is extended approximately exponentially with increasing bias field, and is invariant with changing tweezer depth. The error bars in (c) are similarly given by the square root of the covariance matrix diagonal entry corresponding to the spin depolarization time parameter of the fits. }
\end{figure*}

The experiment is performed at magnetic fields of 1.34 G, in $2.3$-MHz-deep tweezers. These conditions allow us to initialize the qubit in the $\ket{0}$ ground state without the need for large magnetic field changes, by pumping on $\ket{1} \leftrightarrow \ket{F'=3/2, m_{F'}=+1/2}$ transition; however, due to longer pumping times and worse state preparation fidelity, this initialization scheme is not suitable for fast qubit manipulation presented in sec.~\ref{sec:qubit_rotations}. Following the pumping stage, an $X(\pi/2)$ pulse brings the qubits into the superposition of the two ground spin states. After variable Ramsey dark time ($T$), a second $X(\pi/2)$ pulse is applied, and finally the qubit is projected onto one of the spin states through a blow-away measurement as described in Sec.~\ref{sec:qubit_rotations}. Since $X(\pi/2)$ pulses are detuned from resonance by the nuclear-state splitting, the measured $\ket{1}$ population oscillates at the corresponding frequency of $1~\textrm{kHz}$ (Fig.~\ref{fig:coherence}a). 

If no decoherence was present, the oscillations would decay with 1/e time bounded by the lifetime of atoms in tweezers 6.42(3) s (limited by parametric heating). We find that this lifetime is well described by an empirical decay model for the probability of atom survival $P_\mathrm{survival}(T) \propto\mathrm{exp}(-(aT + bT^2))$, with the parameters $a$ and $b$ fixed by a separate lifetime measurement (Fig. \ref{fig:coherence}a). In the presence of noise, e.g. magnetic field instability and nonuniformity across the array, the qubits dephase, causing the contrast of the oscillations to decay faster than the tweezer lifetime bound. In Fig.~\ref{fig:coherence}a, we fit the Ramsey fringes with a cosine of a single frequency and phase, decaying with a Gaussian envelope multiplied by the lifetime decay function. The resultant contrast of the oscillations decays with 1/e time of 3.0(2) s, including both dephasing and the tweezer lifetime, while the extracted $T_2^*$ time characterizing the dephasing alone is equal to 3.7(4) s. This is most likely limited by submilligauss magnetic field fluctuations. 

To further study the coherence properties, we also perform a spin-echo measurement, where an $X(\pi)$ pulse is inserted in the middle of the Ramsey dark time (Fig. \ref{fig:coherence}b). Applying a $Z$ gate for variable time $t$ just before the last $X(\pi/2)$ pulse, allows us to scan over the coherent oscillation of the atomic ensemble (Fig. \ref{fig:coherence}b inset). The extracted contrast of the fringe as a function of dark time $T$ is plotted in Fig. \ref{fig:coherence}b. For certain noise sources such as shot-to-shot magnetic field variation or inhomogeneities across the array, the phases acquired by the atoms during the two dark times cancel each other, suppressing contrast loss. However, mechanisms that cause variation in the qubit frequency on the timescale of the echo arm, such as drifts of the global magnetic field, impose an envelope empirically described by a Gaussian, here with 1/e time equal to $T_2$ = 7.9(4) s. The total contrast loss is given by a product of the Gaussian and the measured lifetime functions, and decays with 1/e time of 4.84($^{+5}_{-3}$) s. The observation of seconds-scale coherence times in megahertz-scale deep traps emphasizes the robustness of this qubit to light-shift-induced dephasing effects.

Another source of decoherence arises due to mechanisms that pump atoms from one qubit state into the other. To experimentally investigate this effect, we prepared the atoms in $\ket{0}$, waited for a variable duration, and then blew away this nuclear-spin state (Appendix \ref{sec:t1}). Any survival is evidence for nuclear-spin depolarization. We examined this effect for a range of trap depths and magnetic fields (Fig. \ref{fig:coherence}c), finding that $T_1$ is invariant as trapping laser intensity changes by a factor of 5 but increases approximately exponentially with increasing magnetic field, and hence with increasing qubit splitting. We note that fluctuations of the transverse magnetic field at kHz-scale frequencies could account for our observations. Under all conditions investigated, we found $T_1 > 10$ s, and that $T_1$ can be extended to the 100 s scale by applying a moderate magnetic field of several gauss. For larger magnetic fields, $T_1$ was found to be too large to measure, given the $5.8(6)$ s 1/e trap loss time scale. The independence of $T_1$ from trap depth suggests that Raman scattering is not a significant contributor. This is consistent with calculations, which give a negligible Raman scattering rate of $<10^{-6}$ s$^{-1}$ for tweezers with a wavelength of 532 nm. For the nuclear qubit, destructive interference in the scattering amplitudes arises in the sum over the hyperfine manifolds of a certain fine structure level. When the detuning from the intermediate states is large in comparison to the hyperfine splitting, the Raman scattering rate is suppressed by many orders of magnitude due to this destructive interference \cite{dorscher2018lattice}. 

\section{\label{sec:sideband} 3d ground-state cooling}

\begin{figure*}
\centering
\includegraphics{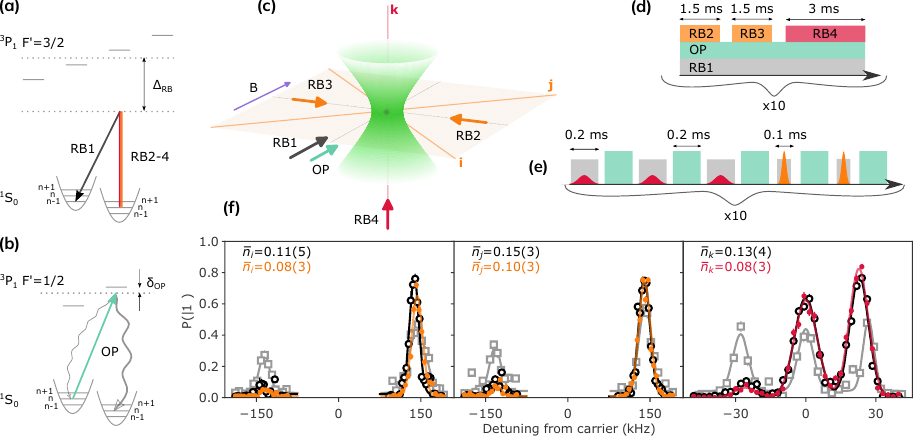}
\caption{\label{fig:sideband} Raman-sideband cooling to the 3D motional ground state. (a) Level diagram for the two-photon transitions employed in cooling. An atom in the $|m_F=+1/2, n\rangle$ state absorbs a photon from the $\pi$-polarized beam (RB2-4) and emits into the $\sigma^+$-polarized beam (RB1). With the relative frequency difference of the two beams tuned a trap frequency away from the carrier, the atom can change its motional state to $n-1$ (red sideband) or $n+1$ (blue sideband), depending on the sign of the difference. The Raman detuning from the intermediate excited state is $\Delta_\mathrm{RB}/2\pi$ = $-183$ MHz. (b) Optical pumping (OP) scheme for energy dissipation during cooling. In the Lamb-Dicke regime, the scattering from the pumping beam is unlikely to alter the atom's motional state but pumps into the opposite spin state, where the cooling can begin anew. We operate with optical pumping $\delta_\mathrm{OP}/2\pi = -2$ MHz red detuned from resonance. (c) Raman beams (RB1-4), optical pumping, and magnetic field ($B$) geometry. RB1 paired with RB2-4 addresses motional states along the $i$, $j$, and $k$ axes respectively. (d) Continuous sideband cooling. RB1 and OP beams are continuously illuminating the atoms, while the remaining RB2-4 are turned on and off iteratively. (e) Pulsed sideband cooling. A pulse of RB1 and one of RB2-4 drives an approximate $\pi$ rotation on the relevant sideband, followed by a step of optical pumping. (f) Sideband thermometry along the $i,~j$ and $k$ axes. Spectra are acquired before sideband cooling (gray, squares), after 3D optimized cooling (black open circles), and after cooling optimized for radial (orange, points) and axial (red, points) directions. Each panel shows the red sideband (left) and blue sideband (right) for the corresponding cooling direction. Axial spectra additionally include the carrier (middle). Mean phonon occupation number ($\bar{n}$) after sideband cooling is quoted for each axis. The error bars correspond to standard deviations of the binomial distributions given by the measured probabilities.
}
\end{figure*}

The $X$ qubit rotations described in the previous sections were performed with copropagating Raman beams. The advantage of this approach is the insensitivity of the qubit operation to the atom's motional state, since the net $k$ vector transferred during the rotation is 0. However, for future applications, such as the promotion of the qubit to the clock state and a Rydberg excitation, where a single beam is employed, motional-state coupling or Doppler shifts will result in reduced pulse fidelity (Appendix \ref{sec:clock_dephasing}). As such, motional-ground-state cooling is desirable for high-fidelity population manipulation within those states. For this reason, we employ Raman-sideband cooling of $^{171}$Yb atoms to near the three-dimensional ground state \cite{monroe1995resolved, kaufman2012cooling, thompson2013coherence}. In the experiments below, we operate with trap frequencies $\{\omega_i/2\pi,~\omega_j/2\pi,~\omega_k/2\pi\} = \{$139.8(8) kHz, 137(1) kHz, 27.4(2) kHz$\}$ along the $i$, $j$, and $k$ axes (Fig. \ref{fig:sideband}c) and $17.1(5)$ MHz tweezer depth (6 mW/trap), in a 10$\times$10 array. 

The atoms are first precooled using the same beams and parameters as employed to load atoms into the tweezers. The subsequent sideband cooling is performed through Raman rotations on the ${^1}\mathrm{S}_0\leftrightarrow{}{^3}\mathrm{P}_1$ $|F'=3/2, m_{F'}=+1/2\rangle$ transition, with a $\sigma^+$-polarized RB1 beam and one of the three $\pi$-polarized RB2-4 beams. The level structure and the beam geometry involved are presented in Fig. \ref{fig:sideband}a-\ref{fig:sideband}c. Pairing RB1 with RB2 or RB3 addresses one of the two radial directions ($i$ and $j$), while pairing RB1 with RB4 is used to address the axial direction ($k$). The relative frequency of the two selected Raman beams is tuned to drive $\ket{m_F=+1/2, n} \leftrightarrow  \ket{m_F=-1/2, n-1}$, while the energy is dissipated by optical pumping on $\ket{F=1/2, m_{F}=-1/2} \leftrightarrow \ket{F'=1/2, m_{F'}=+1/2}$ transition. The Lamb-Dicke parameters for the radial and axial directions are $\eta_{i,j}=0.23$ and $\eta_{k}=0.53$ respectively. Importantly, we find that for the hotter atoms the optical pumping can heat up and eject them from the trap, likely due to differential light shifts. In the nonmagic trapping potentials, the optical pumping beam appears blue detuned for the hotter atoms and can induce sideband heating. If this heating rate exceeds the Raman sideband cooling rate, the atom escapes from the trap. To eliminate this effect, we operate with red-detuned pumping light. This suppresses both the loss and systematic underestimation of the temperature due to the ejection of hot atoms. 

The sideband cooling proceeds in two stages, continuous and pulsed, which are illustrated in Figures~\ref{fig:sideband}d and \ref{fig:sideband}e. The total cooling time is 78 ms. The carrier Raman Rabi frequencies for transitions performed with RB1 and one of RB2-4 are 2$\pi \times$47(2) kHz, 2$\pi \times$48(2) kHz, and 2$\pi \times$13.1(6) kHz respectively. To evaluate performance of the cooling sequence, we carry out sideband spectroscopy with pulses of RB1 and one of RB2-4. Scanning the relative frequency of the two Raman beams, we probe the red (RSB) and blue (BSB) sidebands and extract the mean phonon occupation numbers ($\bar{n}$) from their height ($A_{\mathrm{RSB}}$ and $A_{\mathrm{BSB}}$), according to $\bar{n} = \frac{A_\mathrm{RSB}/A_\mathrm{BSB}}{1-A_\mathrm{RSB}/A_\mathrm{BSB}}$~\cite{monroe1995resolved}. Before sideband cooling, the $\bar{n}$ for each of the axes are $\{\bar{n}_i, \bar{n}_j, \bar{n}_k\} = \{1.6(6), 1.0(4), 4(3)\}$. With the cooling optimized simultaneously in all three dimensions, we achieve $\{\bar{n}_i, \bar{n}_j, \bar{n}_k\} = \{0.11(5), 0.15(3), 0.13(4)\}$ (Figure \ref{fig:sideband}f). Additionally, we can further improve radial temperature by sacrificing ground-state fraction in the axial direction and vice versa. In the first case, we cool the atoms to $\{\bar{n}_i, \bar{n}_j\} = \{0.08(3), 0.10(3)\}$ in the radial directions and in the second case, we cool the axial direction to $\bar{n}_k=0.08(3)$.

In this work, the lowest temperatures obtained with the 3D sideband cooling are influenced by a number of effects. The tweezers exhibit parametric heating, which are measured to be 10(3) quanta/s in the radial direction.  The final temperature can be improved by increasing the cooling rate as well as reducing the tweezer heating rate, the latter of which is largely the result of intensity noise on our trap light source. By increasing the confinement of the weakest (axial) direction of the trap through the addition of a lattice~\cite{young_half_2020}, the three-dimensional cooling rate could be increased substantially. Other factors limiting the cooling rates include increased motional excitation during optical pumping due to the nonmagic trapping and finite lifetime of the excited state, as well as a spread in trapping frequencies due to imperfect AOD tweezer balancing (see Appendix \ref{subsec:balance}). In the future, we will overcome these limitations by performing the Raman sideband cooling in 759 nm spatial-light-modulator-based tweezers. We expect a near-magic angle condition for ${^1}\mathrm{S}_0 \rightarrow {^3}\mathrm{P}_1$ in $^{171}$Yb to exist at this trapping wavelength~\cite{norcia2018microscopic}, while the spatial light modulator will permit improved array homogeneity. As another direction to explore, resolved sideband cooling on the ${^1}\mathrm{S}_0 \rightarrow {^3}\mathrm{P}_0$ transition in the 759 nm lattice has proven successful \cite{brown2017hyperpolarizability} and is a viable option for 3D motional-ground-state preparation in 759 nm tweezers.

\section{\label{sec:conculsion} Conclusion}
In this work, we have demonstrated that $^{171}$Yb tweezer arrays have several salient features --- near-deterministic loading; fast, high-fidelity nuclear qubit control; long nuclear coherence times; and the ability to prepare atoms near the motional ground state via Raman-sideband cooling. Combining scalability and low-entropy preparation with fast, coherent qubit manipulations, this platform will be broadly useful in quantum science applications. 

For connecting the tools developed in this work to metrological and quantum information applications, it will be desirable to develop tweezer arrays at the 759 nm magic wavelength of the clock transition. To maintain favorable features of the 532 nm tweezers demonstrated here, in future work with 759 nm, we will rely on transfer methods already realized with strontium~\cite{young_half_2020, schine2021long}. Once they are loaded into 759 nm tweezers, interactions between $^{171}$Yb nuclear qubits in the ${^1}\mathrm{S}_0$ state can be generated by selectively exciting one nuclear state to ${^3}\mathrm{P}_0$ and driving Rydberg interactions out of that level. Alternatively, rotations of the nuclear spin could be produced directly in the metastable ${^3}\mathrm{P}_0$ level, driving Raman rotations through ${^3}\mathrm{P}_0 \leftrightarrow{}{^3}\mathrm{D}_1$ or ${^3}\mathrm{P}_0 \leftrightarrow{}{^3}\mathrm{S}_1$. With this approach, two-qubit gates could be performed between ${^3}\mathrm{P}_0$ spins with state-selective Rydberg interactions or by mapping one spin of the ${^3}\mathrm{P}_0$ atoms back to the ground state. Beyond quantum information processing, deep, fast circuits implemented in the nuclear-spin qubit could be mapped to the optical clock transition for quantum-enhanced metrology~\cite{pedrozo2020entanglement, marciniak2022optimal}. Furthermore, in 759 nm tweezers, a near-magic angle may exist for the ${^1}\mathrm{S}_0 \leftrightarrow {^3}\mathrm{P}_1$ transition~\cite{norcia2018microscopic}. This could allow for improved imaging fidelity as previously demonstrated in near-magic tweezers with $^{174}$Yb \cite{saskin2019narrow}. Meanwhile, site-resolved shelving utilizing the clock transition could be used to perform local state-preserving qubit measurements, which are a key requirement for most quantum error correction schemes~\cite{terhal2015quantum, fowler2012surface} as well as for clock and entangled clock protocols that fully exploit the intrinsic linewidth of the atoms~\cite{rosenband2013exponential, kessler2014heisenberg, pezze2020heisenberg}. 

\begin{acknowledgments}
We would like to thank Jeff Thompson, Mark Brown, Cindy Regal, Nelson Darkwah Oppong, Andrew Ludlow, and Ben Bloom for helpful discussions. We thank Philipp Preiss, Daniel Slichter, and Aaron Young for close readings of the manuscript. We would also like to acknowledge Mark Brown, Felix Vietmeyer, and Zhenpu Zhang for their contributions to the computer control system. We wish to acknowledge the support of ONR-YIP (N000142012692), ARO (W911NF1910223), AFOSR (FA95501910275), DOE Quantum Systems Accelerator (7565477), NSF (1914534), and NSF Physics Frontier Center (PHY 1734006). This research was supported by an appointment to the Intelligence Community Postdoctoral Research Fellowship Program at JILA at the University of Colorado, Boulder administered by Oak Ridge Institute for Science and Education (ORISE) through an interagency agreement between the U.S. Department of Energy and the Office of the Director of National Intelligence (ODNI). A.S. acknowledges the support from the Funai Overseas Scholarship. W.F.M. acknowledges support of the NIST NRC program. 
\end{acknowledgments}
\hfill \break
\noindent\emph{Note added}.--- Recently, we became aware of complementary work with nuclear-spin qubits of $^{171}$Yb~\cite{ma2021Universal}. 

\appendix

\renewcommand{\thefigure}{A\arabic{figure}}
\setcounter{figure}{0}

\section{\label{sec:experiment_sequence}Experimental sequence}

We use an atomic dispenser as the source of $^{171}\mathrm{Yb}$. The initial trapping and cooling is done with 399 nm light addressing the ${^1}\mathrm{S}_0\rightarrow {^1}\mathrm{P}_1$ transition. Atoms released from the dispenser are slowed by a beam focused onto the emission port of the dispenser and then captured in a 2D magneto-optical trap, which reduces the atoms' velocities in the directions transverse to the science glass cell. The magnetic fields for the slowing beam and 2D MOT are generated by four stacks of permanent magnets arranged around the 2D MOT chamber \cite{lamporesi2013compact, nosske2017two}. The atomic cloud accumulated within the 2D MOT is pushed toward the main science cell with a nearly resonant beam that is chopped at 1 kHz with a 40\% duty cycle. After being pushed through a differential pumping tube, the atoms are trapped in a blue 3D MOT in the main science cell.

\begin{center}
\begin{figure}
    \includegraphics[width=\columnwidth]{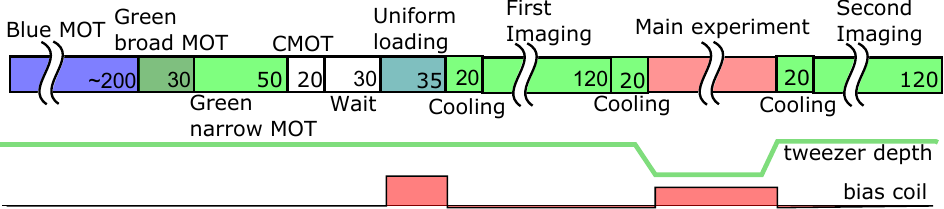}
    \caption{Overview of the experimental sequence in the science cell. The unit of time is milliseconds, and duty cycle is typically less than $1~\mathrm{s}$. The blue MOT uses ${^1}\mathrm{S}_0\leftrightarrow{^1}\mathrm{P}_1$ transition, while the green MOT utilizes ${^1}\mathrm{S}_0\leftrightarrow{^3}\mathrm{P}_1$ narrow-line transition. The compressed MOT (CMOT) increases the average atom number loaded into a tweezer. With the near-deterministic loading scheme, we are left with a single atom in the tweezer $>90$\% of the time, on average. The green cooling beam is on during blue imaging, as well as during the cooling before and after imaging. }
    \label{fig:sequence}
\end{figure}
\end{center}

The following experimental sequence is shown in Fig. \ref{fig:sequence}. The cycle time of the experiment is typically less than 1 s, although the exact timing depends on the experiment performed. Atoms in the 3D blue MOT are transferred to a green MOT that uses the 556 nm ${^1}\mathrm{S}_0\leftrightarrow{^3}\mathrm{P}_1$ narrow-line transition ($\Gamma_\mathrm{g}/(2\pi) =  183 \mathrm{kHz}$). Our green MOT has three steps: broad-line, narrow-line and compression stages. During the first stage, the green laser is artificially broadened by sweeping the detuning from $-40 \Gamma_\mathrm{g}$ to $-7 \Gamma_\mathrm{g}$ at a rate of 50 kHz to increase the velocity capture range. For the second, narrow-line stage the detuning is maintained at $-0.77\Gamma_\mathrm{g}$ and the intensity is ramped down from $157 I_\mathrm{sat}$  to $11 I_\mathrm{sat}$. While we find that at this point, the atoms are sufficiently cold to be loaded into 790-$\mathrm{\upmu K}$-deep tweezer traps, to increase the average number of atoms captured in the tweezers, we employ a final compression stage for the green MOT. Here, the beam intensity is decreased to $1.5 I_\mathrm{sat}$ and the magnetic field gradient is increased from 6 $\mathrm{G/cm}$ to 15 $\mathrm{G/cm}$.

After extinguishing the MOT beams, 30 ms of dead time is incorporated to ensure that any remaining atoms not trapped in the tweezers fall out the tweezer region. We then turn on a 4.9 $\mathrm{G}$ magnetic field parallel to the tweezer polarization and with blue-detuned 556 nm beams propagating in the MOT configuration we achieve single-atom tweezer occupancy at efficiencies in excess of 90\% (see discussion in the main text for more details).

Single atoms trapped in the tweezers are cooled down to $12~\mathrm{\upmu K}$ in 20 $\mathrm{ms}$ by the same green laser beams used for the 3D MOT. However, this intratrap cooling is realized with no magnetic field applied. The green cooling intensity is $10 I_\mathrm{sat}$ and its frequency is $-11\Gamma_\mathrm{g}$ red detuned from the tweezer light-shifted resonance of the ${^1}\mathrm{S}_0\leftrightarrow{^3}\mathrm{P}_1$ $|F'=3/2,~ m_{F'}=\pm1/2\rangle$ transition. The same cooling scheme is employed during imaging.

The main experiment, such as qubit rotations or Raman sideband cooling, is usually sandwiched by two imaging steps (see Sec.~\ref{subsec:image} for detailed discussion). The first image discriminates the traps with loaded atoms, while the second image identifies the tweezer sites that still contain an atom after the experiment is completed. Where indicated in the main text, we lower the tweezer depth during the experiment to suppress differential light shifts or parametric heating from the tweezer intensity noise. In these experiments, a magnetic field is typically applied parallel to the tweezer polarization to define the quantization axis.

\section{\label{sec:experiment_overview}Experimental methods}

\subsection{\label{subsec:Optics}Optics layout around objective lens}

\begin{figure}[tb]
    \includegraphics[width=\columnwidth]{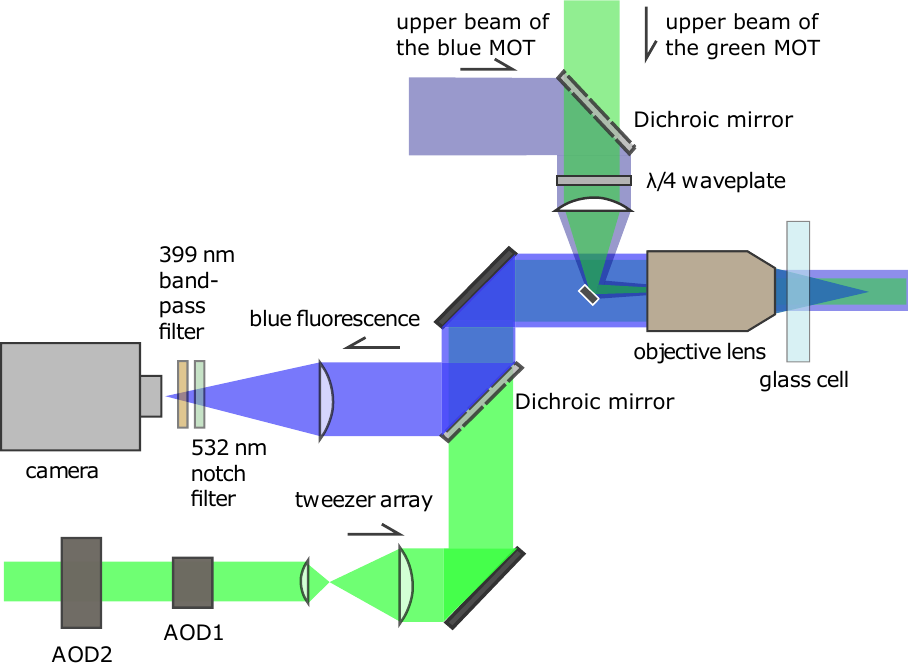}
    \caption{Optics around the objective lens. The $399$ and $556~\mathrm{nm}$ laser beams for the MOTs are focused to the back focal plane of the objective lens and reflected by a mirror, small enough not to degrade the quality of either the tweezers or the image. }
    \label{fig:optics}
\end{figure}

Our optical layout around the objective lens is summarized in Fig. \ref{fig:optics}. Both blue and green vertical upper MOT beams are focused on the back focal plane of the objective lens and reflected by a small, 5-mm-diameter mirror, to ensure collimation inside the science cell. The mirror and its holder are small enough compared to the aperture of the objective lens (32 mm diameter) as not to impact the diffraction-limited imaging and the quality of the tweezer spot.

The tweezer array is formed by the deflection of 532 nm beam from two orthogonal AODs, AOD1 and AOD2, placed in the Fourier plain of the objective. To make tweezer generation more robust, the AODs are spaced with a 4-f lens system (not shown in Fig. \ref{fig:optics}). The beam is then focused through the objective lens to create a 2D tweezer array. The total power in the array is stabilized via an intensity servo actuating on an acousto-optic modulator before the photonic crystal fiber that delivers the light to the setup of Fig. \ref{fig:optics}. The tweezer waist is measured to be $460(24)~ \mathrm{nm}$ by comparing the light shift of the $^3\mathrm{P}_1$ $|F'=3/2,~m_{F'}=\pm1/2\rangle$ state to the power in a single tweezer. The polarizabilities of the $^{171}\mathrm{Yb}$ states required for this measurement were calculated from the values of $^{174}\mathrm{Yb}$ obtained at 532 nm in Ref.~\cite{yamamoto2016ytterbium}. These tweezer waist and light-shift measurements are also consistent with the radial trap frequencies obtained through Raman-sideband spectroscopy (Sec.~\ref{sec:sideband}). 

\subsection{\label{subsec:balance}Tweezer balancing}

\begin{figure*}[tb]
    \includegraphics[width=2\columnwidth]{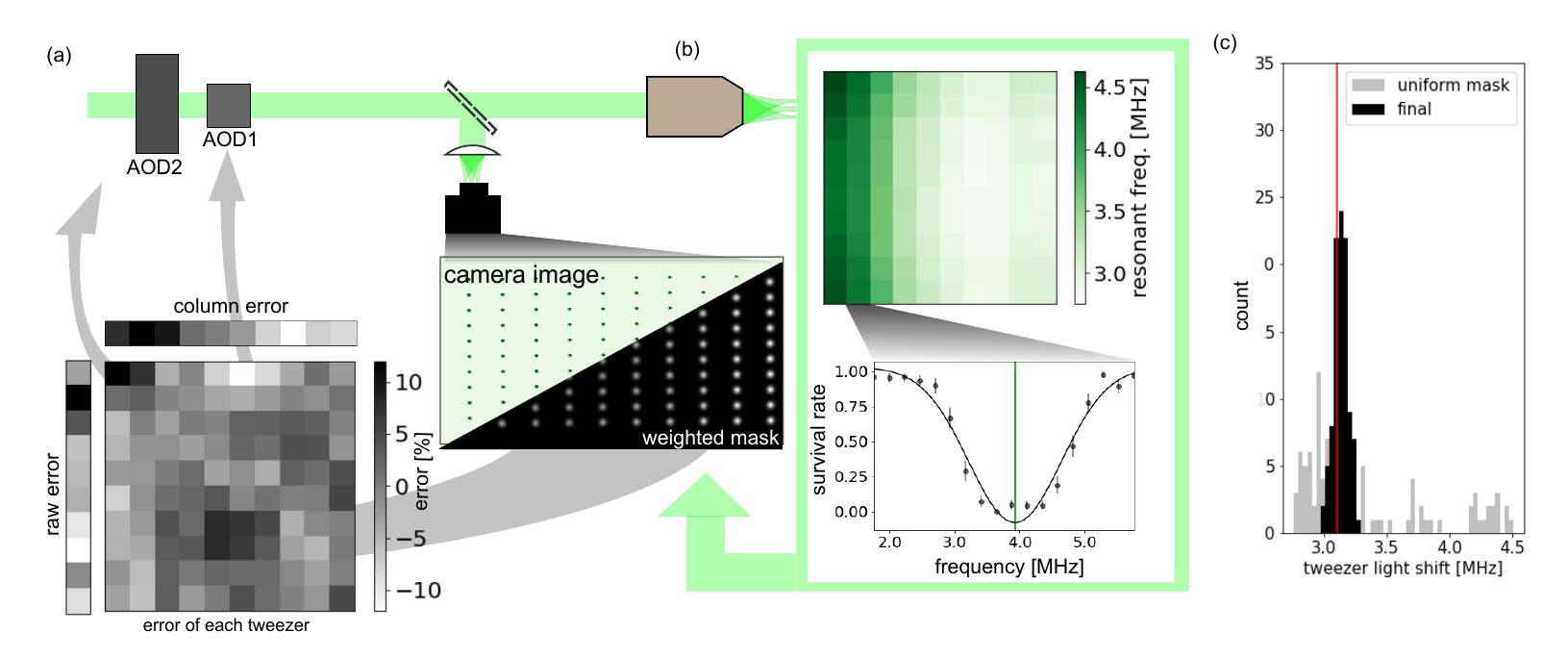}
    \caption{Procedure employed to optimize tweezer balance. (a) Each tweezer's intensity distribution is measured with light picked off before the objective lens and focused onto a camera. The intensity measurement is multiplied by an experimentally determined weighted mask and used to determine the errors for each tweezer. From this 2D array of errors, row and column errors are extracted and proportionally fed back to AOD1 and AOD2, respectively.(b) Example of the spectroscopy result for a uniform mask ($M_{ij}=1$). This information is further utilized to generate a weighted mask. The frequency variable is the red detuning from free-space resonance. (c) Initial (gray) and final (black) distributions of the $^1\mathrm{S}_0\leftrightarrow{^3}\mathrm{P}_1$, $F'=3/2$, $~m_{F'}=1/2$ resonance for the 100 individual tweezers. The red line shows the median of the initial distribution, which is the chosen target value for the balancing.}
    \label{fig:balance}
\end{figure*}

Suppressing tweezer-dependent trap inhomogeneity is critical for many aspects of the experiments presented here. A spread in the tweezer trap depths results in variable differential light shifts on the $^1\mathrm{S}_0\leftrightarrow{^3}\mathrm{P}_1$ transition, which gives rise to the nonuniform cooling performance across the array. Moreover, the variation in the trap frequencies can harm the Raman-sideband cooling efficiency. As discussed in Sec.~\ref{subsec:Optics}, the optical tweezer arrays are produced by two AODs addressing orthogonal directions. We insert $N_1$ and $N_2$ radiofrequency tones into each AOD (typically in this work $N_1=N_2=10$). Those rf tones are generated by a custom-designed  field programmable gate array-based synthesizer, with arbitrary control over the phase and amplitude for each rf tone. Ideally, choosing identical rf amplitudes would generate tweezers of identical intensities. However, practically, this is not the case due to imperfections in our rf and optical systems, leading to intermodulation and nonlinearity. Therefore, we need a protocol for adjusting the RF parameters experimentally. In this section, we discuss our tweezer balancing procedure. Related methods are described in Ref.~\cite{singh2022dual}.

In order to make the tweezer depth as uniform as possible, we balance the tweezer intensity daily with a quick optimization procedure that feeds back on the integrated intensity of each tweezer spot as measured on a camera, subject to occasional calibrations using the atomic signal. Tweezer-site-dependent light shifts found via spectroscopy are used to calibrate a transfer function: from the tweezer intensity measured with the camera to the intensity focused onto the atoms (Fig.~\ref{fig:balance}). This is necessary, since the uniform intensity of the spot array at the camera does not guarantee uniform intensity after the objective lens. The use of the transfer function also ensures efficient daily rebalancing of the array through images acquired with the camera at an intermediate image plane, rather than having to measure and feed back based solely on the atom signal. 

For the tweezer balancing procedure, we fix the phases of the rf tones following the theoretical values \cite{schroeder1970synthesis}, which minimize the variation in total rf power and the chance for the coherent superposition of the multiple rf tones.

Our optimization algorithm starts by taking a picture of the tweezer intensity distribution at an image plane before the objective (Fig. \ref{fig:balance}a). Here, the sum of the counts around the $(i,j)$th tweezer spot $C_{ij}$ is proportional to the power of the corresponding tweezer. Then, we multiply by a weighting factor $M_{ij}$, determined experimentally as discussed below, and obtain the 2D array, $M_{ij}C_{ij}$. We define the balancing error at each tweezer spot $E_{ij}$ as,
\begin{equation}
    E_{ij} = \frac{M_{ij}C_{ij} - \langle MC\rangle}{\langle MC\rangle}
\end{equation}
where $\langle MC\rangle = \frac{1}{N_1N_2}\sum_{ij}M_{ij}C_{ij}$ is the mean value of $M_{ij}C_{ij}$ of the entire array. We convert this 2D error to two 1D errors, corresponding to row and column errors, with two methods. For the row error $E^\mathrm{row}_i$,

\begin{equation}
    \begin{aligned}
        E^\mathrm{row}_i = \begin{cases}
    \frac{1}{N_2}\sum_j E_{ij} & \text{(mean method)} \\
    E_{i,\mathrm{random}(1,...,N_2)} & \text{(random method)}
  \end{cases}
    \end{aligned}
\end{equation}
where $\mathrm{random} ()$ indicates a random integer chosen from among the numbers in the parentheses. We include the ``random method'', because the ``mean method'' often converges to a local minimum where the variation of errors within a row is significant, while the variation among the array is small. Random methods are used for the final optimization following the mean method. The column error is calculated in the same manner. The new rf amplitudes for the vertical and horizontal AODs, $A^{\mathrm{vert}}_\mathrm{new},A^{\mathrm{hor}}_\mathrm{new}$, are calculated using proportional feedback,
\begin{equation}
    A^{\mathrm{hor/vert}}_\mathrm{new} = A^{\mathrm{hor/vert}}_\mathrm{old} - pE^\mathrm{row/column},
\end{equation}
where $p$ is the proportional gain, and $A^{\mathrm{vert}}_\mathrm{old},A^{\mathrm{hor}}_\mathrm{old}$ are arrays containing previous amplitudes.

The weighted mask $M_{ij}$ is generated with a spectroscopy measurement of the site-dependent differential light shifts. Initially, we take $M_{ij}=1$ and balance the tweezers with the method described above. Spectroscopy of the $^1\mathrm{S}_0\leftrightarrow{^3}\mathrm{P}_1 ~\ket{F'=3/2,~m_{F'}=1/2}$ transition performed after such balancing (Fig.~\ref{fig:balance}b) shows $1.75~\mathrm{MHz}~ (10~\Gamma)$ peak-to-peak inhomogeneity on top of a $3.5$ MHz differential light shift. Using the detuning from the free-space resonance at site $(i,j)$, $R_{ij}^{(1)}$, we calculate the mask value as,
\begin{equation}\label{eq:M}
    M_{ij}=M_{ij}^\mathrm{(1)} =\frac{R_{ij}^{(1)}}{R_\mathrm{target}}
\end{equation}
where $R_\mathrm{target}$ is taken as the median of $\{R_{ij}^{(1)}\}$. We find that we can realize a more uniform array when we take the median rather than mean as the target value.

After calibration of the mask weightings, the tweezer balancing follows as before. We typically see a convergence of the total error value to around 10\% peak to peak by several tens of repetition. This could potentially be further improved using more elaborate feedback methods. With the balanced tweezers we take the spectroscopy data again, and obtain a second set of $R_{ij}^{(2)}$ values. From these, we calculate the corresponding $M_{ij}^\mathrm{(2)}$ following Eq.~\ref{eq:M}, and compose a mask according to
\begin{equation}
    M_{ij}=M_{ij}^\mathrm{(1)}\times M_{ij}^\mathrm{(2)}.
\end{equation}

We repeat this procedure several times and finally achieve a distribution of $R_{ij}$ with a standard deviation of $0.05~\mathrm{MHz} ~ (0.3\Gamma)$, or 1.4\% of the magnitude of the differential light shift, which is sufficient for the experiments presented in this paper. The comparison of the initial and final light-shift distribution is shown in Fig.~\ref{fig:balance}(c).

\subsection{\label{subsec:image}Cooling and imaging}

The tweezer array is imaged by collecting photons scattered from a retroreflected low-power beam resonant with the strongly allowed $^1$S$_0\rightarrow ^1$P$_1$ transition at 399 nm. The second-stage ($^1$S$_0\rightarrow ^3$P$_1$) MOT beams are turned on at the same time, to prevent the atoms from being heated out of the array. The survival probability during imaging is optimized by operating with a detuning of -2.04 MHz (11$\Gamma$) from the light-shifted $^3\mathrm{P}_1$ $|F'=3/2,~m_{F'}=\pm1/2\rangle$ resonance and an intensity of 10$I_\mathrm{sat}^g$. To assess the efficiency of cooling under these conditions, we compare release-and-recapture measurements to a Monte Carlo simulation, to extract the atomic temperature \cite{tuchendler2008energy}. The temperature is 12 $\upmu$K when only the cooling light is applied and $\approx$30 $\upmu$K when the imaging beam is also present. We note that both of these temperatures are significantly below the Doppler limit at this detuning, 49 $\upmu$K. Our observations are in reasonable quantitative agreement with a prior measurement, made with a MOT of $^{171}$Yb, which the authors attributed to a sub-Doppler polarization-gradient cooling mechanism obtainable for a transition with a ground state $F>0$, with 3D cooling \cite{maruyama2003investigation}. Furthermore, the atomic temperature is found to scale similarly with increasing 556 nm intensity to the scaling measured by Ref.~\cite{maruyama2003investigation}. We assess the survival and infidelity of the imaging scheme at a range of trap depths, optimizing the cooling parameters at each set point. Our imaging scheme remains effective for trap depths as low as $9$ MHz, half of our operating trap depth, with increasing losses below that level. For the range of trap depths explored, losses are minimized by selecting a detuning of $\approx 60\%$ of the total trap light shift. This dependence suggests that a ``Sisyphus cap" effect arising from a repulsive Sisyphus barrier, such as that observed in Ref.~\cite{cooper_alkaline-earth_2018}, may be playing a role, though further study is needed to fully appreciate the interplay between the sub-Doppler polarization-gradient cooling and the Sisyphus cap mechanisms.

\begin{figure}
    \includegraphics[width=\columnwidth]{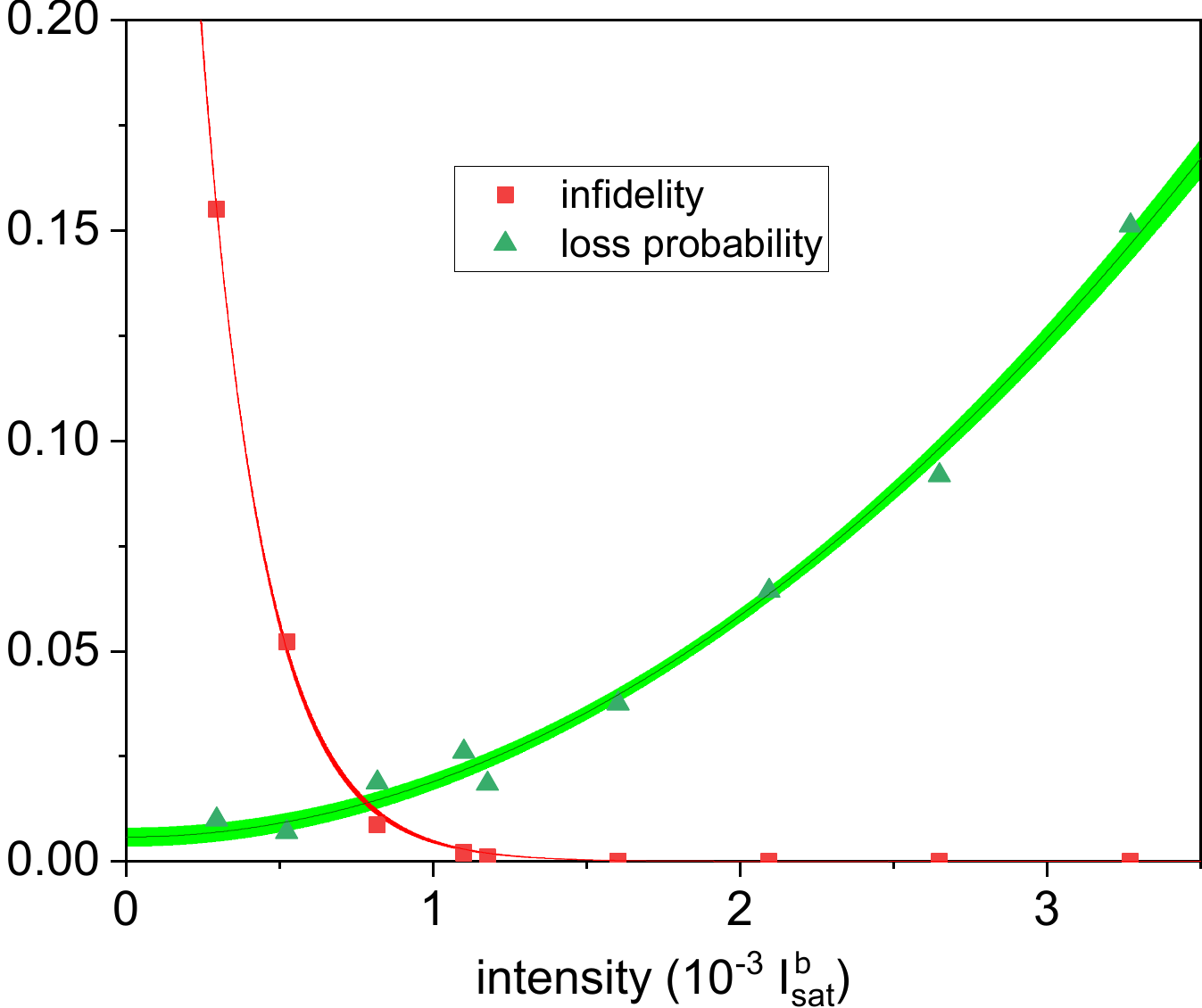}
    \caption{Infidelity and loss probability for an image of duration 120 ms, with variable power in the 399 nm imaging beam. $I_\mathrm{sat}^b = 63$ mW/cm$^2$ is the saturation intensity of the transition at 399 nm. The red curve is an exponential fit to the infidelity data, and the green curve is a quadratic fit to the loss data. Both shaded regions are 1$\upsigma$ confidence intervals.}
    \label{fig:imaging}
\end{figure}

Imaging quality is assessed by fitting a double Gaussian distribution to the photon count histograms (Fig. 1(c)) and setting a threshold between the two distributions that determines whether a given number of photons measured on a pixel will be labeled as corresponding to a singly occupied or to an empty tweezer. We define imaging infidelity as the probability of improperly classifying a single-atom image within the array, i.e., the sum of the area above threshold for the void peak and the area below threshold for the atom peak. The threshold is set so as to minimize the infidelity, defined in this way. Increasing the 399 nm power allows more photons to be collected, decreasing the infidelity of the image. However, this also increases the probability of a given atom to be lost during the course of imaging. Figure \ref{fig:imaging} displays the dependence of infidelity and loss probability on imaging beam power. For this figure, the losses are calculated by using the area under the fitted Gaussian curves as a measure of the atom and void numbers. We find that an imaging intensity of $1.1 \times 10^{-3}$ $  I_\mathrm{sat}^b$ ($I_\mathrm{sat}^b = 63$ mW/cm$^2$ being the saturation intensity of 399 nm transition) represents a reasonable operating condition, with 0.3\% infidelity and a loss probability in the range of 2-3\%. If it is desirable to shorten imaging time, then it is possible to do so by increasing power, leading to increased losses or infidelity. For instance, we find that it is possible to operate with an intensity of $2.3\times 10^{-3}$ $I_\mathrm{sat}^b$ and an imaging time of 63 ms, with similar losses and a modest increase in infidelity.

\begin{figure*}
\includegraphics{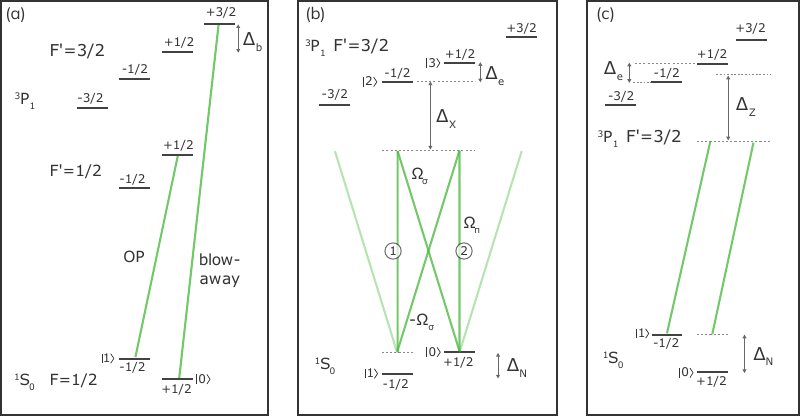}
\caption{\label{fig:qubit_levels} Nuclear-spin preparation, manipulation, and detection. (a) We prepare the spin state in $\ket{0}$ by optical pumping (OP) through ${^3}\mathrm{P}_1$ $\ket{F'=1/2,m_{F'}=+1/2}$ and detect the state destructively by driving the cycling transition $\ket{0} \leftrightarrow \ket{F'=3/2,m_{F'}=+3/2}$ until atoms in the $\ket{0}$ state are expelled from their tweezers. We split the $\ket{F'=3/2,m_{F'}= +1/2}$ and $\ket{F'=3/2,m_{F'}= +3/2}$ states by $\Delta_b/(2\pi)=35$ MHz during blow-away detection to minimize off-resonant pumping through the $\ket{F'=3/2,m_{F'}=+1/2}$ level by the blow-away beam. (b) $X$ rotations in the high-Rabi regime. The two Raman pathways couple the states $\ket{0}$ and $\ket{1}$ through two different excited states $\ket{2}$ and $\ket{3}$. The splitting between the nuclear-spin states is $\Delta_N$, determined by both the external magnetic field and light shifts from the drive beam. The detuning from the excited states is $\Delta_X/(2\pi)\simeq -180$ MHz, much larger than the $\Delta_e/(2\pi) = 2.65$ MHz splitting between the $\ket{F'=3/2,m_{F'}=\pm1/2}$ levels. The Rabi coupling in the different arms is set by the dipole coupling matrix elements and the polarization of the drive beam. (c) Level diagram showing the $\sigma+$-polarized $Z$ beam coupling to the $\ket{F'=3/2,m_{F'}=+1/2}$ and $\ket{F'=3/2,m_{F'}=+3/2}$ states. In the case of $\Delta_Z \gg \Delta_e$, the larger coupling matrix element on the stretched transition to $\ket{F'=3/2,m_{F'}=+3/2}$ gives a larger light shift on the $\ket{0}$ spin level. This splits the nuclear-spin states and allows us to perform fast rotations about the $Z$ axis. }
\end{figure*} 

Comparisons between the measured loss rates for our imaging system using our Andor SCMOS Marana and the preliminary imaging fidelity measurements using an Andor EMCCD iXon camera indicate that we may improve our imaging performance significantly with an EMCCD. At an EMCCD imaging infidelity of 0.03\%, about an order of magnitude lower than our operating SCMOS infidelity, we expect that we can use 40\% of our operating 399 nm imaging intensity. Under these conditions, we extrapolate that the atom loss probability will be about 0.7\% per image. Alternatively, targeting a higher imaging infidelity at the 0.6\% level that we measure for a 60 ms image, we expect to be able to shorten the imaging time to 25 ms with loss rates $\leq 1\%$ per image.

\section{\label{sec:SPAM}Loss correction}
The dominant source of state preparation and measurement error is atom loss (2-3.5\%), with the majority of this loss occurring during the imaging steps. Blow-away detection and imaging infidelity errors are smaller, typically $0.5$-$0.7\%$ and $0.2$-$0.3\%$ respectively. For subfigures of Fig. \ref{fig:qubit_rotations} with plotted probabilities $P(\ket{1})$, we associate the detection of an atom in the second image to the $\ket{1}$ state, normalizing the resulting probability of detecting $\ket{1}$ by the measured atom loss probability (without correcting for imaging infidelity). The probability of atom survival without the blow-away detection pulse is measured for a given experiment as $P(s)$ and the normalized probability of the atoms being in the $\ket{1}$ state $(\ket{m_F=-1/2})$ is given by $P(\ket{1}) = P(a)/P(s)$ where $P(a)$ is the probability that an atom is imaged after the blow-away detection pulse is applied. The probability of atoms being in the $\ket{0}$ state $(m_F=+1/2)$ is then $1-P(\ket{1})$. The uncertainties in these probabilities are propagated from the uncertainties of the $P(a)$ and $P(s)$ measurements and are both 
\[
\sigma_{0,1} = \frac{P(a)}{P(s)} \sqrt{\left(\frac{\sigma_{a}}{P(a)} \right)^2 + \left(\frac{\sigma_{s}}{P(s)} \right)^2}.
\]
The randomized benchmarking data are not loss corrected. The success probabilities of the benchmarking sequences are the raw values set by the measured atom and void detection probabilities where an atom is associated with target state $\ket{1}$ and a void is associated with target state $\ket{0}$.

\section{\label{sec:qubit}Nuclear qubit preparation, rotation, and read-out}

Figure \ref{fig:qubit_levels} shows diagrams of the beams and levels used for preparation, manipulation, and readout of the nuclear-spin state. We prepare the ${^1}\mathrm{S}_0$ spin state in $\ket{m_F=+1/2}\equiv\ket{0}$ by optical pumping through the ${^3}\mathrm{P}_1$ $\ket{F'=1/2,m_{F'}=+1/2}$ state with a $\sigma_+$-polarized beam. For spin detection, we destructively read out the spin state by blowing away atoms in the $\ket{0}$ state with a beam that is resonant with the cycling transition to the ${^3}\mathrm{P}_1$ $\ket{F'=3/2, m_{F'}=+3/2}$ state. The blow-away beam heats $\ket{0}$ atoms out of the tweezers on a few hundred microsecond timescale. Typically, the blow-away beam is applied for several ms with the exact time chosen as a trade-off between errors caused by the slow pumping from $\ket{m_F=+1/2}\equiv\ket{1}$ to $\ket{0}$ through the excited state $\ket{F'=3/2,m_{F'}=+1/2}$ and errors caused by the small probability of an atom in $\ket{0}$ surviving the destructive pulse. We typically obtain a combined detection error around $6 \times10^{-3}$ when minimizing the sum of these two error sources. 

As discussed in the main text, in the strong-Rabi regime we drive Raman transitions between the nuclear-spin states using a single beam with polarization components along both the atom plane normal and the quantization axis. This beam couples the spin states through two different pathways as shown in Fig.~\ref{fig:qubit_levels}b. For the beam geometry used here, the strength of the two circular components is equal $\Omega_{\sigma}\equiv|\Omega_{\sigma+}|=|\Omega_{\sigma-}|$. The arms that couple the ground-state spin to $\ket{F'=3/2,m_{F'}=\pm3/2}$ do not drive transitions but they do lead to differential light shifts of the qubit states due to the different detunings from these excited states. This effect changes the qubit splitting from its value due to the magnetic bias field alone and, in the strong-drive regime, the splitting goes from $\Delta_N/(2\pi) = -1.25$ kHz to $+54.2$ kHz during an $X(\pi/2)$ pulse. To describe the two-pathway Raman transitions, we focus on the four $\pm 1/2$ states $\ket{0}$, $\ket{1}$, $\ket{2}$, and $\ket{3}$ (Fig.~\ref{fig:qubit_levels}b), and absorb the light shifts due to the $\pm 3/2$ states into the parameters $\Delta_X$ and $\Delta_N$. In a rotating frame given by
\[
U(t) = e^{i (\omega_0-\Delta_X)t\ket{3}\bra{3} +i(\omega_0-\Delta_X)t\ket{2}\bra{2}},
\]
with $\omega_0$ the optical frequency of the $\ket{1}\leftrightarrow\ket{2}$ transition, the Hamiltonian describing these four levels and single drive beam is
\[
H=\begin{pmatrix}
\Delta_N       & 0 & \frac{1}{2}\Omega_{02} & \frac{1}{2}\Omega_{03}\\[0.6em]
0              & 0 & \frac{1}{2}\Omega_{12} & \frac{1}{2}\Omega_{13}\\[0.6em]
\frac{1}{2}\Omega^*_{02}  & \frac{1}{2}\Omega^*_{12} & \Delta_X & 0\\[0.6em]
\frac{1}{2}\Omega^*_{03}  & \frac{1}{2}\Omega^*_{13} & 0 & \Delta_X  + \Delta_e
\end{pmatrix}.
\]
For the beam geometry used here, we define $\hat{y}$ to point along the atom plane normal and $\hat{z}$ to point along the quantization axis. Decomposing electric field of the $X$ beam $\mathbf{E} = E_y \text{cos}(\omega t + \phi_{H/V})\hat{y} + E_z \text{cos}(\omega t)\hat{z}$ into a spherical tensor basis and taking the positive rotating component \cite{brink, devanathan, steck}, we find
\[
\mathbf{E^{(+)}} = -i \frac{E_y}{2\sqrt{2}}e^{i(\omega t + \phi_{H/V})}(\hat{e}^*_{-1} + \hat{e}^*_{+1})  + \frac{E_z}{2} e^{i\omega t} \hat{e}^*_0
\]
This field gives coupling terms
\[
\Omega_{02/13} = \frac{i e^{i\phi_{H/V}}}{\sqrt{6}} \left( \Gamma_0 \frac{3\pi \epsilon_0 \hbar c^3}{\omega_0^3} \right)^{1/2} \frac{E_y}{\hbar}
\]
for transition decay rate $\Gamma_0$ and transition frequency $\omega_0$ \cite{steck}. Defining $\Omega_\sigma$ and $\Omega_\pi$ as the magnitudes of $\Omega_{02/13}$ and $\Omega_{03/12}$ respectively,
\[
H=
\begin{pmatrix}
\Delta_N  & 0 & \frac{1}{2} e^{i\phi}\Omega_\sigma & \frac{1}{2}\Omega_\pi\\[0.6em]
0 & 0 & \frac{1}{2}\Omega_\pi & \frac{1}{2} e^{i\phi}\Omega_\sigma \\[0.6em]
\frac{1}{2} e^{-i\phi}\Omega_\sigma & \frac{1}{2}\Omega_\pi & \Delta_X & 0 \\[0.6em]
\frac{1}{2}\Omega_\pi & \frac{1}{2} e^{-i\phi}\Omega_\sigma & 0 & \Delta_X + \Delta_e\\
\end{pmatrix}
\]
where $\phi$ is the phase between the $\pi$ and $\sigma$ single-photon Rabi frequency components. Note that this phase is different than the phase between the $\hat{y}$ and $\hat{z}$ electric field components $\phi_{H/V}$ and the two phases are related by $\phi_{H/V} = \phi - \pi/2$. Adiabatic elimination of the excited levels leads to
\begin{widetext}
\[
H_{eff} = -\frac{1}{4}\begin{pmatrix}
 -4\Delta_N +\left( \frac{1}{\Delta_X}\Omega_\sigma^2 + \frac{1}{\Delta_X + \Delta_e}\Omega_\pi^2  \right)
 & \left(  \frac{e^{i\phi}}{\Delta_X} \Omega_\pi \Omega_\sigma + \frac{e^{-i\phi}}{\Delta_X + \Delta_e} \Omega_\pi \Omega_\sigma \right) \\[0.6em]
 \left(  \frac{e^{-i\phi}}{\Delta_X} \Omega_\pi \Omega_\sigma + \frac{e^{i\phi}}{\Delta_X + \Delta_e} \Omega_\pi \Omega_\sigma \right) 
 &  \left( \frac{1}{\Delta_X}\Omega_\pi^2 + \frac{1}{\Delta_X + \Delta_e}\Omega_\sigma^2  \right) 

\end{pmatrix}.
\]
\end{widetext}

In our experiment, $\Delta_X \gg \Delta_e$, and neglecting the excited state splitting gives a form of the Hamiltonian where the importance of the phase between the polarization components is clear,
\[
H_{eff} \simeq -\frac{1}{4\Delta_X}\begin{pmatrix}
 -4\Delta_N \Delta_X + \Omega_\pi^2 + \Omega_\sigma^2 
 & 2~ \text{cos}(\phi) \Omega_\pi \Omega_\sigma \\[0.6em]
2~ \text{cos}(\phi) \Omega_\pi \Omega_\sigma 
& \Omega_\pi^2 + \Omega_\sigma^2
\end{pmatrix}
\]
and the approximate Raman Rabi frequency is given by $\Omega_X \simeq \text{cos}(\phi)\Omega_\pi\Omega_\sigma/\Delta_X = \text{sin}(-\phi_{H/V})\Omega_\pi\Omega_\sigma/\Delta_X$. This implies a maximum Raman Rabi frequency for $\phi=0,\pi$ or $\phi_{H/V}=\pm \pi/2$, corresponding to a circularly polarized beam. The ratio of $|E_z|/|E_y|=0.54$ gives relative coupling strengths $\Omega_\sigma = 0.9\Omega_\pi$. At our detunings of $177$--$184$ MHz, we obtain Raman Rabi frequencies up to $\Omega_X/(2\pi)=1.77$ MHz with $<40$ mW of power addressing the atom array.

In the strong-driving regime, control of the rotation axis is complicated by the Raman Rabi frequency dependence on the phase $\phi$. As is clear above, at a relative phase of $\phi=\pi/2$ the Raman Rabi becomes small. Also the stabilization of this phase is difficult when using two separate driving beams. To control the rotations around the $Z$ axis, we instead add another beam running along the quantization axis with $\sigma+$ polarization as shown in Fig. \ref{fig:qubit_levels}c. This beam similarly has a large detuning compared to the ${^3}\mathrm{P}_1$ linewidth, $\Delta_Z/(2\pi) = -164$ MHz and couples $\ket{1}$ to $\ket{F'=3/2,m_{F'}=+1/2}$ and $\ket{0}$ to $\ket{F'=3/2,m_{F'}=+3/2}$. The larger coupling matrix element of the $\ket{0}\leftrightarrow \ket{F'=3/2,m_{F'}=+3/2}$ transition gives a correspondingly larger light shift on the $\ket{0}$ spin state. This beam splits the qubit states with a larger frequency than can be achieved easily with an external field. As shown in the main text, we obtain splittings $\Delta_N/(2\pi)= -0.77$ MHz using a total of 11 mW to address the entire atom array.

\subsection*{\label{subsec:qubit_errors}Qubit gate errors}
To investigate the sources of errors in our nuclear-spin rotations, we estimate the error rates due to Raman scattering by our $X$ and $Z$ beams, measured intensity noise, and the uncompensated precession of our $X$ axis. Then, we discuss prospects for minimizing these errors in future experiments.

\begin{figure}
\includegraphics{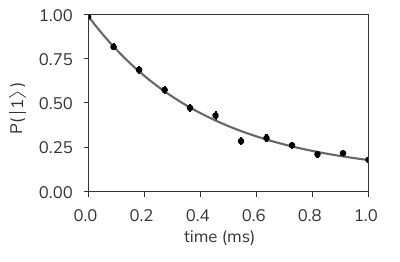}
\caption{\label{fig:Z_scattering} Estimating the scattering rate of the $Z$ beam. The spins are prepared in the $\ket{1}$ state and the $Z$ beam is turned on for the time shown. Scattering causes the spin to relax to a population set by the branching ratios from the two excited levels that the $Z$ beam scatters from: $\ket{F'=3/2,m_{F'}=+1/2}$ and $\ket{F'=3/2,m_{F'}=+3/2}$. The decay timescale is $0.43(4)$ ms, consistent with the scattering rate estimated from the observed differential light shift produced by this beam. The error bars correspond to standard deviations of the binomial distributions given by the measured probabilities.}
\end{figure} 

The Raman scattering rates of the qubit rotation beams are estimated from a depolarization measurement as well as calculations based on the measured $X$ and $Z$ oscillation frequencies. Figure \ref{fig:Z_scattering} shows a measurement of spin depolarization in the presence of the $Z$ beam. The spin is prepared in the $\ket{1}$ state and the $Z$ is turned on for the time range shown. The decay time constant is $0.43 \pm 0.04$ ms, consistent with the calculated scattering for this level given below. 

The excited state splittings and the nuclear-spin state splitting are small relative to the beam detuning. Under these assumptions, the measured Rabi frequency ($\Omega_X/(2\pi) =  1.47$ MHz) of our $X (\pi/2)$ pulses gives the strength of the driving field and this field strength can then be used to calculate the Raman scattering and Rayliegh scattering rates (see below). Calculating the rates as in Ref.~\cite{uys2010decoherence} for the $X$ beam gives a total decoherence rate of an equal spin superposition of $5.6\times 10^3$ s$^{-1}$. Comparing to the time of an $X (\pi/2)$ pulse, this corresponds to a gate error rate $\leq 1.0\times 10^{-3}$. For the $Z$ beam, this equal superposition decoherence rate is estimated to be $2.3\times 10^3$ s$^{-1}$ and corresponds to a $Z (\pi/2)$ gate error of $9\times 10^{-4}$ in a typical $Z$ gate.

Intensity noise on the $X$ and $Z$ pulses is another substantial source of gate error. We measure both short-term (shot-to-shot) and long-term (experiment-to-experiment) fluctuations in the $X$ pulse area on a fast photodiode and find fractional standard deviations of 0.01 and 0.03, respectively. For our $X$ Raman drive, the rotation angle is proportional to pulse area and the resulting fractional angle standard deviation $\sigma_\theta$ due these intensity fluctuations on a single $\pi/2$ rotation is 1.2$^\circ$ and 2.8$^\circ$, respectively. Calculating the corresponding phase flip error rates as $\text{sin}^2(\sigma_\theta \sqrt{2/\pi})$ gives gate errors of $3\times 10^{-4}$ and $1.6\times 10^{-3}$.

In this work, the $X$ gate also has a unitary error due to the detuning of the drive in the high-Rabi case. For the single drive beam, this detuning is equal to the splitting of the nuclear-spin states as seen in Fig.~\ref{fig:qubit_levels}b. This splitting is small, $2\pi\times 1.25$ kHz, at the bias field we operate at, but the $X$ beam also causes these states to split due to different detunings from the excited levels. The differential light shift on the nuclear-spin states is $2\pi\times 55.5$ kHz but of opposite sign to the splitting coming from our bias field, so that the total splitting and thus detuning is estimated to be $2\pi\times 54.2$ kHz. In the time of a single $\pi/2$ pulse, this corresponds to a precession about the $Z$ axis of 3.3$^\circ$ and a gate error of $2.1\times 10^{-3}$. Note that there is also some precession about the $Z$ axis in the time between gates, but for our typical gap time of 500 ns this only contributes $1.5\times 10^{-5}$ to the total error of a single $\pi/2$ gate. Our calibration of $Z (\pi/2)$ gates has this free precession naturally built in.

Finally, we observe some variation in the driven $X$ and $Z$ oscillation frequencies across the atom array. However, these inhomogeneities contribute to the global $\pi/2$ rotation errors at a level $<10^{-5}$.

For future experiments based on these spin control techniques, it is useful to consider how these error rates can be improved and up to what limits. Starting with intensity noise, improvements in AO stability or sample and hold techniques should help significantly. For example, demonstrated 0.2\% fractional intensity errors for similar pulse lengths \cite{de2018analysis} would correspond to an error rate of $<10^{-6}$. It should also be possible to suppress unitary detuning errors significantly. One option is to make $X$ gates out of composite $X$ and $Z$ rotations that are tailored to correct for the detuning \cite{cummins2001resonance}. Another is to drive these $X$ rotations using the $F'=1/2$ manifold instead of the $F'=3/2$ levels used here, which gives a much smaller differential splitting of the nuclear spin as the shifts nearly cancel for the case of equal Rabi coupling strength in all Raman arms. However, the most straightforward method to minimize the $X$ detuning error may be to increase the beam detuning from $F=3/2$ and increase the power. This has the added benefit that it would also suppress scattering errors.

Scattering errors are caused both by Raman scattering, resulting in a spin flip, as well as elastic Rayleigh scattering, resulting in phase shifts between the two spin states. As shown in Ref.~\cite{uys2010decoherence}, the two types of errors have different dependence on the scattering amplitudes, with the error rate caused by the former given by the standard Kramers-Heisenberg formula,
\[
\Gamma_{ij} = \Omega_R^2 \Gamma_0 \sum_q \left(\sum_{F',m_F'} A_{F',m_F',q}^{i,j} \right)^2
\]
and the error rate caused by the latter given by a sum over scattering amplitudes of the two spin states,
\[
\Gamma_{el} = \Omega_R^2 \Gamma_0 \sum_q \left(\sum_{F',m_F'} A_{F',m_F',q}^{+1/2,+1/2}-A_{F',m_F',q}^{-1/2,-1/2} \right)^2
\]
where $\Omega_R = \mu E_0/2\hbar$ and $\mu = \left( \Gamma_0 3\pi \epsilon_0 \hbar c^3/\omega_0^3 \right)^{1/2}$
with transition frequency $\omega_0$, electric field $\bar{E}=E_0\sum_q b_q \hat{\epsilon}_q$, natural linewidth $\Gamma_0$, and scattering amplitudes from spin state $i$ to spin state $j$ given by
\[
A_{F',m_F',q}^{i,j} = \frac{b_q \bra{j} \bar{d}\cdot \hat{\epsilon}^*_{q+(i-j)} \ket{F',m_F'}\bra{F',m_F'}\bar{d}\cdot \hat{\epsilon}_{q}\ket{i}}{\Delta_{F',m_F'} \mu^2}
\] for detuning $\Delta_{F',m_F'}$ from the intermediate state $\ket{F',m_F'}$. In that work, the effect of the two errors are quantified by the combined rate at which they cause decoherence of an equal superposition of spin states, $\Gamma_d=\left( \Gamma_{-1/2,+1/2}+\Gamma_{+1/2,-1/2}+\Gamma_{el}\right)/2$. For an optimized $X$ beam polarization, the ratio of this decoherence rate to Rabi rate $\Omega_r$ is \[
\frac{\Gamma_d}{\Omega_r} \simeq \frac{\Gamma_0}{\sqrt{6}\Delta_0}\left|\frac{\Delta_{hf}}{\Delta_0 - \Delta_{hf}}\right|\text{.}
\]
where $\Delta_{hf}$ is the hyperfine splitting of $^3P_1$ and $\Delta_0$ is the detuning from the $F'=3/2$ hyperfine states, assumed to be much larger than the splitting of different $m_F'$ states.
From this expression it can be seen that the error rate for a given rotation angle is limited by the available laser power and desired Rabi rate. 

In future work, it may be preferable to use a qubit defined by nuclear-spin states of the ${^3}\mathrm{P}_0$ level, and drive Rydberg interactions directly between the atoms in this long-lived clock state. In that case, similar Raman rotations of the spins could be driven using the 1388 nm ${^3}\mathrm{P}_0 \leftrightarrow {^3}\mathrm{D}_1$ transition. This transition is in a telecom band, addressable with commercially available high-power lasers, and is of relevance for architectures combining tweezers with silicon waveguides \cite{covey2019telecom}. Alternatively, it would be possible to drive Raman rotations using 649 nm ${^3}\mathrm{P}_0 \leftrightarrow {^3}\mathrm{S}_1$ transition. For us, this transition has the advantage of being readily available to utilize for local qubit addressing, when paired with high-NA objective, diffraction limited at 649 nm, and single-beam Raman rotation approach.

\section{\label{sec:t1}$T_1$ measurements}

To examine the depolarization of the atomic sample, atoms were prepared in $\ket{0}$ through optical pumping. After a variable delay, the atoms still in $\ket{0}$ were blown away, and then the remaining population of atoms in $\ket{1}$ were detected. A physical shutter is used to fully extinguish the qubit beams during the delay, preventing leak-through scattering from coupling the spin states. For the results reported in the main text, the qubit population as a function of time was modeled with the following differential equations, subject to the initial condition, $n_{\ket{1}}(t=0) = p$:

\[
    \dot{n}_{\ket{0(1)}} (t) = \frac{n_{\ket{1(0)}} (t) - n_{\ket{0(1)}}(t)}{T_1} - (a+2bt)n_{\ket{0(1)}} (t).
\]

\begin{figure}
\includegraphics[width = 0.5\textwidth]{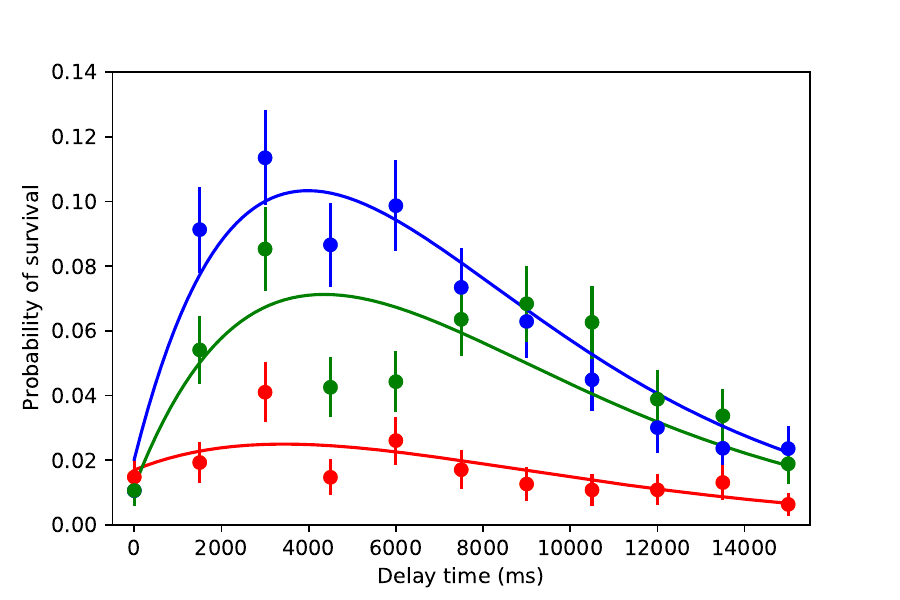}
    \caption{\label{fig:t1rawdata} By first preparing the atoms in $\ket{0}$, waiting a variable delay, and then blowing away that spin state, the depolarization time constant $T_1$ can be measured. The survival probability is plotted as a function of delay time for three different magnetic field conditions: $B\approx 0$ G (blue), 0.7 G (green), and 2 G (red). The lines represent the fit applied to the data. The fits yield values for $T_1$ of 17.0(1.5) s for the blue points, 26.7(4.8) s for the green points, and 119(29) s for the red points. Error bars represent 1$\upsigma$ uncertainty of the binomial distributions given by the measured probabilities.}
\end{figure}

Here, $T_1$ is the depolarization time constant, and $a$ and $b$ are parameters characterizing loss from the trap. Note that solving the differential equation for the total atom number, $P_\mathrm{survival}(t)=n_{\ket{0}}(t) + n_{\ket{1}}(t)$ yields the heuristic loss equation from the main text, $P_\mathrm{survival}(t) \propto\mathrm{exp}[-(at + bt^2)]$. In these fits, the values of $a$ and $b$ are fixed through an independent measurement of the decrease of total atom number as a function of delay time, $P_\mathrm{survival}(t)$, and the statistical uncertainty of the measured values of $a$ and $b$ did not limit the uncertainty of $T_1$. Since parametric heating varies with trap depth, $a$ and $b$ were measured separately at each trap depth investigated. The fits employ $T_1$ and the initial spin polarization purity $p$ as free parameters. Figure \ref{fig:t1rawdata} shows three curves with representative data, with $T_1 = 17.0(1.5)$, 26.7(4.8), and 119(29) s. The relatively high error bars for these measurements are attributable to the fact that the 1/e trap loss timescale, $5.8(6)$ s for these measurements, is substantially shorter than $T_1$ and tends to reduce the magnitude of the signal. Furthermore, we found that $T_1$ varied significantly, by approximately a factor of 4, from day to day with otherwise identical conditions, suggesting that depolarization arises from time-varying environmental noise. To check the consistency of the model, a further sequence of measurements was made, in which a $\uppi$ pulse before the blow-away allowed the detection of atoms remaining in $\ket{0}$. This series of measurements was taken over the same range of magnetic fields as the measurements reported in Fig. \ref{fig:coherence}c, and reasonable statistical agreement was found between the two methods of measuring $T_1$.

The differential equations listed above make the assumption that the rate of pumping from $\ket{0}$ to $\ket{1}$ is the same as the rate from $\ket{1}$ to $\ket{0}$. To test this assumption, a further measurement was performed in which a $\uppi$ pulse was used prior to the delay, to test the rate of depolarization for the $\ket{1} \rightarrow \ket{0}$ channel. The value of $T_1$ measured through this method was found to agree at the $<1\upsigma$ level with that measured through atoms instantiated in $\ket{0}$, though it is only possible to statistically verify that these rates are equal at the $\approx$ 20\% level.

\section{\label{sec:clock_dephasing}Clock pulse motional dephasing}

\begin{figure}
\includegraphics{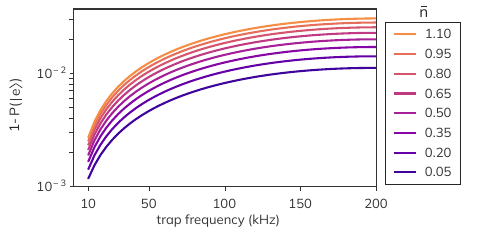}
\caption{\label{fig:clock_dephasing} Clock pulse motional dephasing. The calculated infidelity of the population transfer to the clock state $\ket{e}$ is plotted as a function of trap frequency and mean motional occupation number $\bar{n}$. The Rabi frequency is fixed at $\Omega_c/(2\pi)=200$ kHz. The trap is assumed to be magic, with equal trap frequencies in the ground and clock states.}
\end{figure} 

The clock transition ${^1}\mathrm{S}_0 \leftrightarrow{}{^3}\mathrm{P}_0$ will play a central role in future work on Yb tweezer arrays, both in metrological applications and also as a metastable level from which we will produce Rydberg interactions between atoms. This means that it will be important to perform high-fidelity $\pi$ pulses that map atoms to and from the clock state. One limitation to this fidelity is the dephasing of a Rabi drive that results from motional-state dependence of coupling terms between the ground and clock state. This motivates Raman-sideband cooling to the motional ground state in any experiments involving clock manipulations. Here we consider the effect of temperature on clock pulse fidelity and estimate the motional dephasing limits to clock $\pi$-pulse fidelities. In a rotating frame defined by a beam with detuning $\delta$ from the excited state, and Rabi frequency $\Omega_c$, we use a 1D Hamiltonian describing the atom-field interaction and motional energies of a harmonically trapped atom along the axis of beam propagation,
\begin{widetext}
\[
H_c=\left(-\delta\ket{e}\bra{e}\right) \otimes \mathbbm{1}_M +  \frac{\Omega_c}{2}\left( \sigma \otimes e^{-i \eta (a+a^\dagger)} + \sigma^\dagger \otimes e^{i \eta (a+a^\dagger)}  \right) + \mathbbm{1}_S \otimes \omega \sum_{n=0}^{N'} \left(n+\frac{1}{2}\right)\ket{n}\bra{n}
\]
\end{widetext}
where $\ket{g}$ and $\ket{e}$ are the ground and clock states, $\mathbbm{1}_{M}$ and $\mathbbm{1}_{S}$ are identities on the motional and spin spaces respectively, $\eta$ is the Lamb-Dicke parameter, $a$ and $a^\dagger$ are the motional annihilation and creation operators, $\omega$ is the trap frequency, and we cut off the sum over motional states at $N'$. For atoms in magic wavelength tweezers at 759 nm, the trap frequencies are the same in the ground and clock states, $\omega_e = \omega_g = \omega$. We assume this magic condition and a Rabi frequency of $\Omega_c/(2\pi)=200$ kHz, and calculate the maximum population transferred to the clock state with a resonant $\pi$ pulse at a range of temperatures given by the mean occupation number $\bar{n}$ and over a range of trap frequencies; see Fig. \ref{fig:clock_dephasing}. The calculation selects the $\pi$-pulse time that maximizes the population transfer at each temperature and trap frequency, and uses motional levels up to $N'=7$ and detuning $\delta=0$.

Figure \ref{fig:clock_dephasing} highlights the importance of sideband cooling to future applications involving the clock state. The temperature dependence is twofold, since cooling to lower $\bar{n}$ will allow for operation at lower trap frequencies without loss, at the same time as reducing motional dephasing directly. The motional dephasing effects shown here can likely be improved using composite pulses or adiabatic rapid passage, but we still expect the temperature to play some role in determining the final fidelity of population transfer.

\bibliography{theBib}

\end{document}